\documentclass[12pt]{article}
\usepackage{a4wide}
\usepackage{graphicx}
\usepackage{amssymb}
\usepackage{amsmath}
\usepackage{slashed}
\usepackage{cite}
\usepackage{multicol}
\usepackage{braket}
\usepackage{bmpsize}
\usepackage{xcolor}

\newcommand{\be}{\begin{equation}}
\newcommand{\ee}{\end{equation}}
\newcommand{\ba}{\begin{eqnarray}}
\newcommand{\ea}{\end{eqnarray}}

\begin{document}

%\twocolumn
\allowdisplaybreaks

\begin{titlepage}
\begin{flushright}
\end{flushright}
\vfill
\begin{center}
{\Huge Loop quantum effects on direct detection prediction in two-scalar dark matter scenario}
\vfill
{\large Karim Ghorbani}\\
{{\it Physics Department, Faculty of Science, Arak University, Arak 38156-8-8349, Iran}}\\[1cm]

{\large Parsa  Ghorbani}\\
{\it {Physics Department, Faculty of Science, Ferdowsi University of Mashhad, Iran}}
\end{center}
\vfill
\begin{abstract}
We investigate the effect of quantum corrections on the elastic 
scattering cross section of dark matter off  nucleus in 
two-scalar dark matter model. 
Among two extra singlet scalars in the two-scalar model, the lighter one is stable and plays the role of dark matter candidate and the heavier one contributes in dark matter co-annihilation processes in thermal history of the early universe.  It is already known that the two-scalar model at tree level, unlike the single-scalar dark matter model, can easily evade the bounds from direct detection (DD) experiments. The claim here is that taking into account the loop effects, in some regions of the parameter space, the DM-nucleon cross section becomes larger than the tree level contribution. Therefore, loop effects move the regions which were below the neutrino floor at tree level, up to the regions which are detectable by future DD experiments.  

\end{abstract}
\vfill
%keywords: Beyond the standard model, dark matter experiments, dark matter theory
%{\bf PACS:11.30.Rd, % Chiral symmetries
%          12.39.Fe, % Chiral Lagrangians
%          13.25.Jx, % Decays of other mesons
%          14.40.Aq. % pi, K, and eta mesons
% }
\vfill
%\footnoterule
{\footnotesize\noindent }

\end{titlepage}

\section{Introduction}
\label{int}
A well-known natural scenario for dark matter (DM) is the thermal production of weakly interacting massive particles (WIMPs) in the early universe \cite{Lee:1977ua,Steigman:1984ac,Bergstrom:2000pn,Steigman:2012nb,Leane:2018kjk,Arcadi:2017kky}. 
We may decipher the particle nature of dark matter by its interaction with
normal matter in direct detection (DD) experiments.
However, the DM interaction with nucleons may be so weak such that its cross section resides below the {\it neutrino floor} (NF) and thus not detectable 
in the current DD experiments.
On the theoretical side, we generally compute the DM-nucleon cross section
at tree level in perturbation theory while ignoring the presumably small higher order corrections. 

There are noticeably two classes of models with dark matter candidates escaping the strong bounds from DD experiments. One scenario is that the scattering cross section of DM-nucleon tends to zero at tree level due to some symmetry breaking structure, for example, a pseudo-Goldstone boson as DM candidate in the complex scalar model with softly broken symmetry
\cite{Gross:2017dan,Gonderinger:2012rd,Barger:2010yn}. As such, in models with scale symmetry breaking, the DD cross section at tree level may be 
reduced significantly at some regions in the parameter space \cite{Ghorbani:2022muk, Ghorbani:2024twk}.
The second avenue deals with models wherein the DM-nucleon scattering cross section is momentum or velocity suppressed, giving rise to DM candidates evading DD bounds easily. 
Among models of this type we can recall thermal DM candidates which interact 
with nucleons through pseudoscalar operators
\cite{Ghorbani:2014qpa, Berlin:2015wwa,Fan:2015sza,Yang:2016wrl,Baek:2017vzd,Ghorbani:2017jls,DiazSaez:2021pmg,Chen:2024njd}.

In case we find viable regions in the parameter space where DM-nucleon cross section is small 
or suppressed at tree level, it deems unnecessary to add quantum loop corrections to the DM-nucleon scattering cross section. Because it is commonplace that loop corrections are subleading in perturbation theory. However, in some models it is proved otherwise.
Therefore, by incorporating quantum corrections, regions with very small DM-nucleon cross sections and not accessible by DM direct detection experiments, 
may shift above the neutrino floor and become exploratory regions by 
the present or future DD experiments. 
There are a large number of works in this direction 
with results indicating that the quantum corrections generally alter the viable parameter space considerably \cite{Li:2018qip,Herrero-Garcia:2018koq,Hisano:2018bpz,Han:2018gej,Azevedo:2018exj,Ishiwata:2018sdi,Ghorbani:2018pjh,Ertas:2019dew,Li:2019fnn,Chao:2019lhb,Glaus:2019itb,Borschensky:2020olr,Bell:2018zra,Abe:2018emu,Cho:2023hek}. 

Relevant to our purpose, we recall two minimal extensions of the Standard Model 
where almost all part of the parameter space is excluded by the constraints 
from the observed relic density and direct detection bounds: the singlet scalar 
dark matter model \cite{McDonald:1993ex} and 
the singlet fermionic dark matter model \cite{Kim:2008pp}.
In these two models the same coupling appears in both annihilation cross section and 
direct detection cross section. It is found that no regions can be found to respect 
both the DD bound and the observed relic density.  

Now, we introduce another avenue where quantum loop effects
in DM-nucleon cross section turn out to be prominent in regions not 
excluded by the current direct detection experiments or in regions below the neutrino floor. 
In the pertinent models, the coupling entering the DM-nucleon scattering 
cross section at tree level, can have a little efficacy on the DM annihilation cross section, 
and in fact there are other couplings which play a major role on determining the 
the DM relic abundance. Two such models as next to minimal extension of 
the Standard Model are two-scalar model \cite{Ghorbani:2014gka} and 
two-fermion model \cite{Ghorbani:2018hjs}.
It is shown that the DM-nucleon cross section in two-fermion model is 
subject to sizable loop corrections in some regions of interest \cite{Maleki:2022zuw}.

In this research we take the two-scalar dark matter model where the light 
scalar is stable and becomes the DM candidate.  
The phenomenology of this model with DM-nucleon scattering at tree level 
is carried out in \cite{Ghorbani:2014gka}, finding a large viable parameter
space respecting the observed relic density and DD bounds. 
In addition, there exist regions with viable parameters locating below the neutrino floor. 
The question is that how DM-nucleon cross section at one loop level may affect regions 
below the neutrino floor and regions below the DD bounds respecting other constraints.   
Generally, there exist regions already below the neutrino floor (DD upper bounds) which 
may go up after including the quantum loop effects.
Within the same model, the coscattering effects are studied in \cite{DiazSaez:2024nrq}.
Different scenarios for two-scalar model with emphasis on dark matter phenomenology
are investigated recently in \cite{Habibolahi:2022rcd,Basak:2021tnj,DiazSaez:2021pfw,Ghorbani:2019itr}. 

The paper consists of the following parts.
The DM model with two singlet scalars interacting with the SM Higgs is describe in sec.~\ref{model}.
As well, bounded from below conditions and the Higgs invisible decay upper bound are discussed.
In sec.~\ref{Tree-update} we consider the relic density and dark matter scattering cross section off the nucleus  
by providing numerical results while updating the bounds from the latest direct detection experiments.  
In sec.~\ref{LQC} we introduce the leading quantum corrections 
appearing in triangle and box Feynman diagrams and provide the 
effective scattering amplitude for the elastic scattering of 
dark matter off the nucleus. 
Our main results including the loop effects for the DM-nucleon cross section 
are given in sec.~\ref{results}. 
Finally we finish with conclusion in sec.\ref{conclusion}. 
In addition, we provide the DM annihilation cross sections, DM-nucleon cross section at tree level, 
and loop functions for the box diagram in Appendices $A$, $B$ and $C$, respectively.

\section{Two-Scalar Model}
\label{model}
We recount the two-scalar model as a renormalizable extension of the SM possessing two gauge singlet scalars $\varphi_1$ and $\varphi_2$ under SM gauge symmetries. 
The two scalars are connected to the SM particles via the SM Higgs. 
We apply a ${{\mathbb{Z}}_{2}}$ symmetry under which the two singlet scalars transform
as $\varphi_1 \to -\varphi_1$ and $\varphi_2 \to -\varphi_2$ . The relevant potential including the SM Higgs and two extra singlet scalars in its minimal form is written as 
\begin{equation}
 {\cal V}(H, \varphi_1, \varphi_2) = \mu _{H}^2 H^\dagger H +  \lambda_H (H^\dagger H)^2 + (\alpha_1 \varphi_1^2 + \alpha_2 \varphi_2^2 + 2 \alpha_{12}
 \varphi_1 \varphi_2) H^\dagger H.
 \end{equation}
It is assumed that the two scalars take a zero vacuum expectation value (VEV), while the SM Higgs doublet is expanded around its VEV in the unitary gauge as 
\begin{equation}
\label{2}
\qquad      H= \frac{1}{\sqrt{2}}\begin{pmatrix} 0  \\ v+h \\ \end{pmatrix},
\end{equation}
with $v=246$ GeV being the Higgs' VEV.
The interaction term $\sim \varphi_1 \varphi_2 H^\dagger H$  induces off-diagonal mass terms for the two neutral scalars $\varphi_1$ and $\varphi_2$,
in the mass matrix,
\begin{equation}\label{10}
{{M}^{2}}=\left( \begin{matrix}
m_{\varphi_1}^2               &   \alpha_{12} v^2  \\
 \alpha_{12} v^2  &   m_{\varphi_2}^2   \\
\end{matrix} \right) \,.
\end{equation}

This enforces a rotation in the space of the singlet scalars to transform the 
scalar fields into their mass eigenstates. By introducing the mass mixing angle, $\epsilon$, we take a rotation as 
\begin{equation}
 \phi_1 = \varphi_1 \sin \epsilon + \varphi_2 \cos \epsilon \,, ~~~
 \phi_2 = \varphi_1 \cos \epsilon - \varphi_2 \sin \epsilon  \,,
\end{equation}
and introduce the physical fields $\phi_1$ and $\phi_2$ with physical masses
$m_1$ and $m_2$, respectively. As laid out in \cite{Ghorbani:2014gka}, 
the coupling, $\alpha_{12}$ can now be obtained in terms of the scalar masses and the mass mixing angle, $\epsilon$,
\begin{equation}
 \alpha_{12} = \frac{2 \sin 2\epsilon}{v^2} (m_1^2-m_2^2) \,.
\end{equation}
We take the scalar field $\phi_2$ with smaller mass $m_2$ as our DM candidate. 
We note that the interaction terms $\lambda_1 \varphi_1^4$, $\lambda_2 \varphi_2^4$, 
$\lambda_{12} \varphi_1^2 \varphi_2^2$, $\lambda_{13} \varphi_1 \varphi_2^3$ and $\lambda_{31} \varphi_1^3 \varphi_2$, do not contribute to the DM annihilation cross section as well as in the DM-nucleon cross section at tree level. 
However, in case we pick negative values for all the couplings $\alpha_1$, $\alpha_2$ and 
$\alpha_{12}$, to be assured of vacuum stability, we may set the couplings of the above five interaction terms at fix values 
other than zero. Since our plan in this research is to include the leading quantum corrections we will see that 
these five interaction terms do not play a role. 
Therefore in this model we are left with five independent free parameters: 
$m_1, m_2, \alpha_1, \alpha_2$ and $\alpha_{12}$. The scalar mass difference defined as $\delta = m_1 - m_2 >0$ is used in our computation. In the following sections we may use the identity 
$m_2 \equiv m_\text{DM}$. 

As a requirement for the vacuum stability, the potential part of the Lagrangian should fulfil the bounded from below condition. The relevant formulas are found in \cite{Habibolahi:2022rcd} for a generic potential with two scalars: 
\begin{equation}
 (\Delta > 0~~\text{and}~~A > 0)~~~\text{or}~~~(\Delta > 0~~\text{and}~~B >0),  
\end{equation}
where 
\begin{equation}
\begin{split}
&A = 8ac-b^2\\
&B = 64a^3e-16a^2c^2+16ab^2c-16a^2bd-3b^4\\
&\Delta = 256 a^3e^3 - 192 a^2bde^2 - 128 a^2c^2e^2 + 144a^2cd^2e -27a^2d^4+144 ab^2ce^2-6ab^2d^2e\\
 & - 80 abc^2de + 18abcd^3 + 16ac^4d^2 
+16ac^3d^2 - 27 b^4 e^2 + 18b^3cde -4b^3d^3 - 4 b^2c^3e + b^2 c^2d^2.
\end{split}
\end{equation}
Since in the present study we have set, 
$\lambda_1, \lambda_2, \lambda_{12},\lambda_{13}$ and $\lambda_{31}  =0$, 
the parameters $a,b,c,d,e$ suited to our conditions read
\begin{equation}
 a =  - \frac{\alpha_1^2}{4\lambda_H},~b = -\frac{\alpha_1 \alpha_{12}}{\lambda_H},~c= -\frac{2\alpha_{12}^2+\alpha_1 \alpha_2}{2\lambda_H},
 ~d= -\frac{\alpha_2 \alpha_{12}}{\lambda_H} ,~e = - \frac{\alpha_2^2}{4\lambda_H}. 
\end{equation}

Moreover, in this model the SM Higgs can decay invisibly in three different ways as: 
$h \to \phi_1 \phi_1$, $h \to \phi_2 \phi_2$ and $h \to \phi_1 \phi_2$, if it is allowed kinematically. Thus the total decay width will be modified as 
\begin{equation}
\begin{split}
\Gamma^{\text{tot}}_\text{higgs} &= \cos^2(\epsilon)~\Gamma^{\text{SM}}_\text{higgs} 
+ \Theta(m_h -2m_1) \Gamma(h\to \phi_1 \phi_1) + \Theta(m_h -2m_2) \Gamma(h\to \phi_2 \phi_2) \\&
+ \Theta(m_h -m_1-m_2) \Gamma(h\to \phi_1 \phi_2) \,,
\end{split}
\end{equation}
where $\Gamma^{\text{SM}}_{higgs}$ is the Higgs decay width in the SM, 
$\Theta$ is the step function and $m_h \sim 125$ GeV is the Higgs mass. 
The decay width of the Higgs in three different channels are provided by

\begin{equation}
 \Gamma(h\to \phi_1 \phi_1) = \frac{(\alpha_1 \sin^2 \epsilon+ \alpha_2 \cos^2 \epsilon + 2 \alpha_{12} \sin \epsilon \cos \epsilon)^2v^2}{8\pi m_h} \sqrt{1-4m_1^2/m_h^2} \,,
\end{equation}

\begin{equation}
 \Gamma(h\to \phi_2 \phi_2) = \frac{(\alpha_1 \cos^2 \epsilon+\alpha_2 \sin^2 \epsilon  -2\alpha_{12} \sin \epsilon \cos \epsilon)^2v^2}{8\pi m_h} \sqrt{1-4m_2^2/m_h^2} \,,
\end{equation}

\begin{equation}
 \Gamma(h\to \phi_1 \phi_2) = \frac{((\alpha_1-\alpha_2) \sin \epsilon \cos \epsilon + \alpha_{12} \cos 2\epsilon)^2 v^2}{8\pi m_h^3} \sqrt{m_h^2-(m_1+m_2)^2}  
 \sqrt{m_h^2-(m_1-m_2)^2} \,.
\end{equation}
The experimental upper limit at $95\%$ CL is found on the invisible Higgs decay, such that Br($h\to$ invisibles) $\lesssim 0.18$ \cite{CMS:2018yfx}. This latter constraint becomes more effective for small mass of the scalars or large mixing angle. 

It is also necessary to estimate the decay width of the heavier 
scalar. When the mass difference of the two scalars is smaller than the 
Higgs mass, $\delta < m_h$, then the decay $\phi_1 \to \phi_2 \bar f f$ is mediated by an off-shell Higgs particle. The SM fermion is indicated by $f$. 
The decay width is obtained as 
\begin{equation}
 \Gamma(\phi_1 \to \phi_2 \bar f f) = \frac{3 m_f^2 N_c[(\alpha_1-\alpha_2) \sin \epsilon \cos \epsilon + \alpha_{12} \cos 2\epsilon]^2}{128 \pi^3 m_1^3} 
 \int \int dt~du \frac{(t-4m_f^2-m_2^2+m_h^2)}{(t-m_h^2)^2+\Gamma_h^2 m_h^2}
 \,,
\end{equation}
where $t$ and $u$ are the Mandelstam variables in the relevant decay kinematics, and $N_c$ is the number of color for the SM fermion. 
The decay life-time of the heavy scalar is $\tau = \Gamma^{-1}$. 
We apply the code {\tt CalcHEP} \cite{Belyaev:2012qa} to evaluate the decay width numerically. 
The scalar life-time for a set of parameters with reasonable magnitudes,  
$\delta = 10$ GeV, $m_1 = 300$ GeV and $\alpha_1, \alpha_2, \alpha_{12} \sim 0.5$,
is estimated as $\tau \sim 10^{-9}$ sec. It is also found that for larger mass
of the heavy scalar, the life-time does not change significantly.    
Furthermore, we have verified that for larger values of $\delta$, the decay 
life-time becomes smaller because of the larger available phase space. 
The life-time of the heavy scalar is very smaller than the age of the universe
in the regions of the parameter space which is relevant in this study.

\section{Annihilation Cross Section and DM-nucleon cross section at Tree Level}
\label{Tree-update}

\begin{figure}
\hspace{.7cm}
\begin{minipage}{.6\textwidth}
\includegraphics[width=.82\textwidth,angle =0]{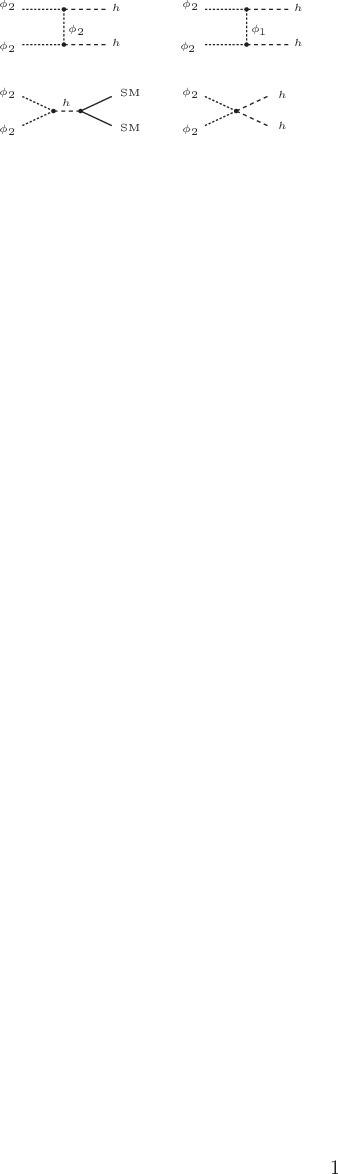}
\end{minipage}
\hspace{1.2cm}
\begin{minipage}{.40\textwidth}
\includegraphics[width=.34\textwidth,angle =0]{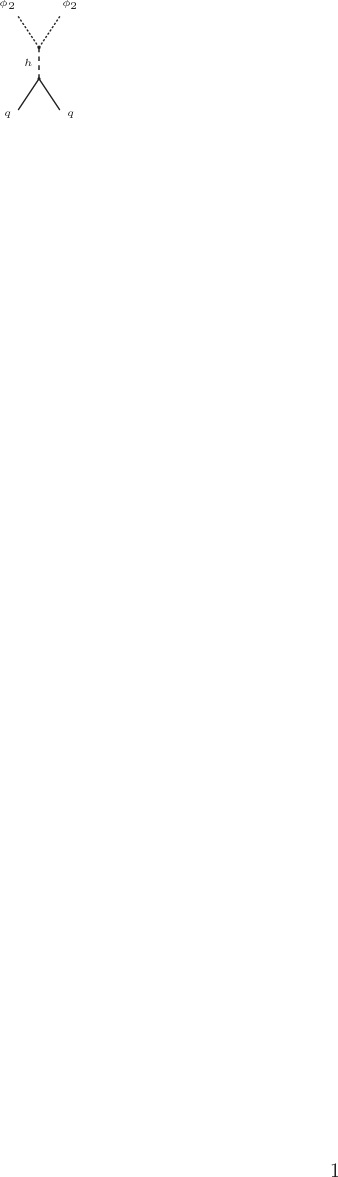}
\end{minipage}
\caption{In the left panel Feynman diagrams for the DM annihilation are shown with only two particles in the final state. The diagram for the DM-nucleon elastic scattering is shown in the right panel.} 
\label{Annihilation-DD}
\end{figure}
The aim of this section is to reanalyze and update what is found within the present model in \cite{Ghorbani:2014gka} from the calculations of the relic density and the DM-nucleon cross section at {\it tree level}. 
The DM density today depends on the so-called freeze-out temperature in the early universe. Around the freeze-out temperature the Hubble expansion rate exceeds the DM annihilation rate and on the other hand, the kinetic energy of the SM particles is low enough which leads to the suppression of the DM production. At this epoch the DM particles become non-relativisitc and go out of equilibrium and thus the DM density remains unchanged afterwards.
The time evolution of the DM number density depends on the (co)annihilation processes
of the two WIMPs. In the following we outline these (co)annihilation processes.   

There are three channels through which the scalar DM can annihilate.
1) Through $s$-channel by Higgs particle as a mediator; with the SM fermions, gauge bosons and the Higgs in the final state. 2) Through $t$- and $u$-channel by one of the scalars as mediator; with a pair of the Higgs particles in the final state. 3) Annihilation into a pair of Higgs particles via a contact interaction. In the left panel of Fig.~\ref{Annihilation-DD}, the corresponding Feynman diagrams for the DM annihilation are shown.
We have not shown diagrams with more than two particles in the final state (i.e., $\phi_2 \phi_2 \to hhh$).

In addition, co-annihilation diagrams will be obtained by simply replacing one $\phi_2$ by one $\phi_1$ in the initial states. 
If we replace the two $\phi_2$ fields with $\phi_1$ fields, 
the annihilation diagrams for the heavier scalar will be obtained.   
In a model with two WIMPs, in principle one should solve two coupled Boltzmann 
equations to obtain the time evolution of the number density of each scalar. 
However, in practice it will be sufficient to solve a 
single Boltzmann equation (sum of the two coupled equations) 
by considering effective annihilation and co-annihilation cross sections 
\cite{Griest:1990kh,Edsjo:1997bg}. In the sum, terms describing the conversion processes and terms describing the decay processes cancel 
each other \cite{Edsjo:1997bg}.  
Therefore the sum of the number densities of the two scalars, $n=n_1+n_2$, 
will change via (co)annihilation of the two scalars, whose governing 
Boltzmann equation reads,
\begin{equation}
 \frac{dn}{dt} + 3 H n = \langle \sigma_{\text{eff}} v \rangle (n_{eq}^2 - n^2) \,,
\end{equation}
where $\langle \sigma_{\text{eff}} v \rangle$ is the 
the thermal average over effective cross section times 
the relative velocity of DM particles at temperature $T$. 
The effective cross section is given by the following expression,
\begin{equation}
 \sigma_{\text{eff}}= g_{\text{eff}}^{-1} \Big[ \sigma^{22}_{\text{ann}} + \sigma^{11}_{\text{ann}} (1+ \frac{\delta}{m_2})^3 e^{-2\delta/T} 
 +2 \sigma^{12}_{\text{co-ann}} (1+ \frac{\delta}{m_2})^{3/2} e^{-\delta/T} 
 \Big]  \,,
\end{equation}
where $\sigma^{22}_{\text{ann}}$, $\sigma^{11}_{\text{ann}}$ and 
$\sigma^{12}_{\text{co-ann}}$, indicate the annihilation cross section 
of DM, annihilation cross section of heavy scalar and co-annihilation cross section, respectively. The effective degrees of freedom is, 
$g_{\text{eff}} = 1 + (1+\delta/m_2)^{3/2} e^{-\delta/T}$.

Let us assume for a moment that $\alpha_{12} = 0$ (thus $\epsilon = 0$). 
This will turn the model back into its simplest form, i.e. only one singlet scalar in the model. 
In this simple case the annihilation cross section contains terms each of them
proportional to $\alpha_2^2$, $\alpha_2^3$ or $\alpha_2^4$. 
Thus the DM relic density is inversely proportional to these couplings.
Now looking at the relevant vertex one can easily see that the DM-nucleon cross section 
depends on the same coupling, $\alpha_2$. 
In the right panel of Fig.~\ref{Annihilation-DD} the diagram for the DM-nucleon elastic scattering is depicted. 
Since the DM-nucleon cross section and DM relic density dependency on the couplings go in opposite way, when varying the coupling $\alpha_2$, as it is shown in \cite{Ghorbani:2014gka}, it is not possible to respect both the DD bounds and the observed relic density at the same time (except for a resonance region where $m_2 \sim m_h/2$).   

Now set $\alpha_{12} \ne 0$, and we get back to our present model. 
This brings in new interactions which give pivotal contribution to the DM annihilation cross section.
We first identify three couplings related to the scalar-Higgs interactions. 
The strength of the interaction vertex $\phi_2 \phi_2 h$ is equal to $v\kappa_{22}$ with 
$\kappa_{22} = \alpha_1 \cos^2 \epsilon + \alpha_2 \sin^2 \epsilon -2 \alpha_{12} \cos \epsilon \sin \epsilon$, 
the strength of the interaction vertex $\phi_1 \phi_2 h$ is equal to $v\kappa_{12}$
with $\kappa_{12} = \alpha_{12} \cos 2\epsilon + \cos \epsilon \sin \epsilon (\alpha_1 - \alpha_2)$, 
and finally the strength of the interaction vertex $\phi_1 \phi_1 h$ is $v \kappa_{11}$ with 
$\kappa_{11}= \alpha_1 \sin^2 \epsilon + \alpha_2 \cos^2 \epsilon + 2 \alpha_{12} \cos \epsilon \sin \epsilon$.
Now the DM-nucleon cross section at tree level is only dependent on the coupling $\kappa_{22}$ while 
the DM annihilation cross section depends on the three couplings $\kappa_{11}$, $\kappa_{22}$ and $\kappa_{12}$. As a result, new viable regions in the parameter space might show up. 
The reason hinges on the fact that by incorporating these new contributions 
it becomes feasible to get small DM-nucleon cross section (small $\kappa_{22}$) 
respecting the current DD bounds, and in order to get the DM relic density 
right one can regulate the couplings $\kappa_{12}$ and $\kappa_{11}$ accordingly.
In Appendix A, the annihilation cross sections are given for the full model.
The DM-nucleon elastic scattering cross section at tree level is given in Appendix B.

\begin{figure}
\hspace{-.5cm}
\begin{minipage}{.52\textwidth}
\includegraphics[width=.71\textwidth,angle =-90]{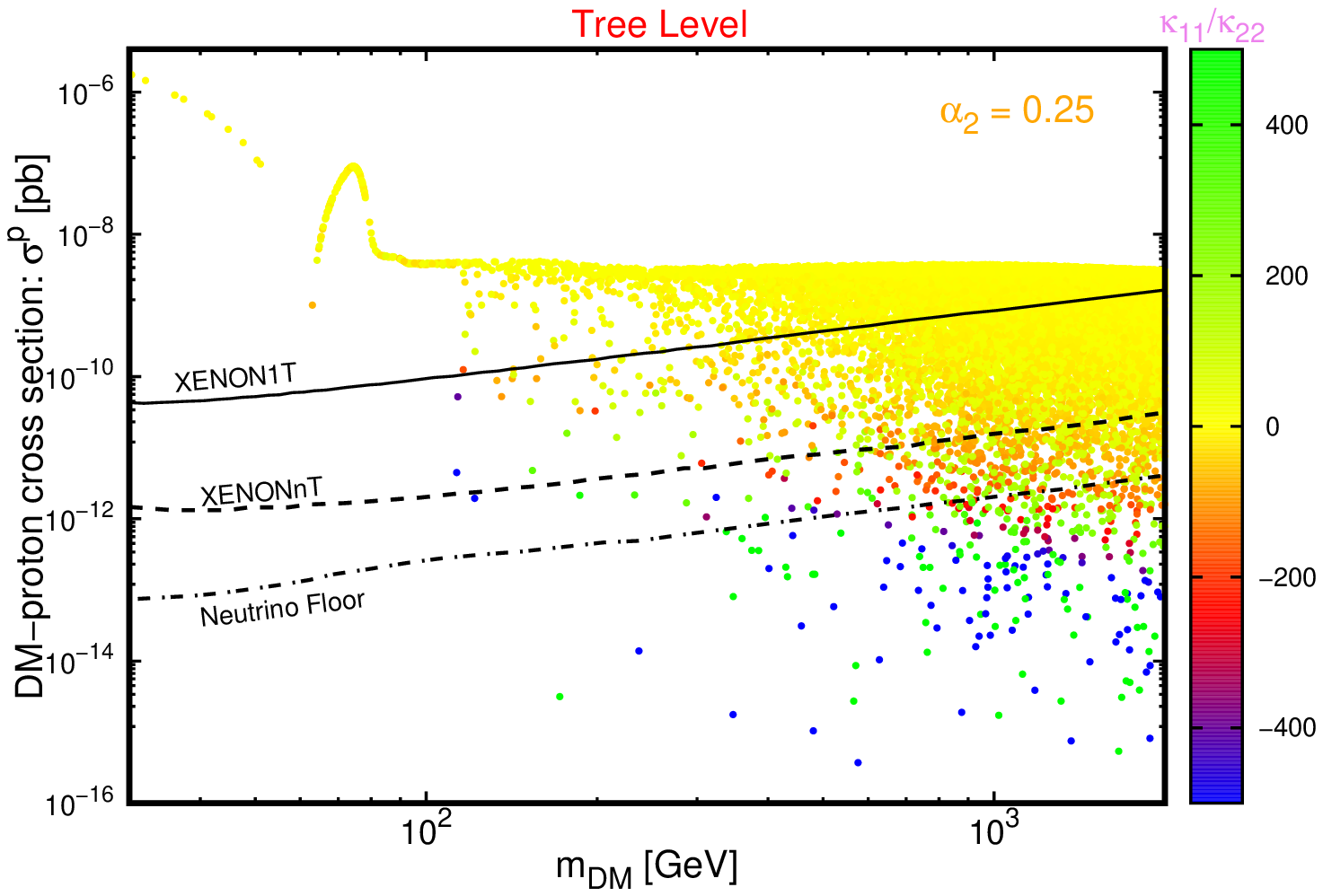}
\end{minipage}
\begin{minipage}{.52\textwidth}
\includegraphics[width=.71\textwidth,angle =-90]{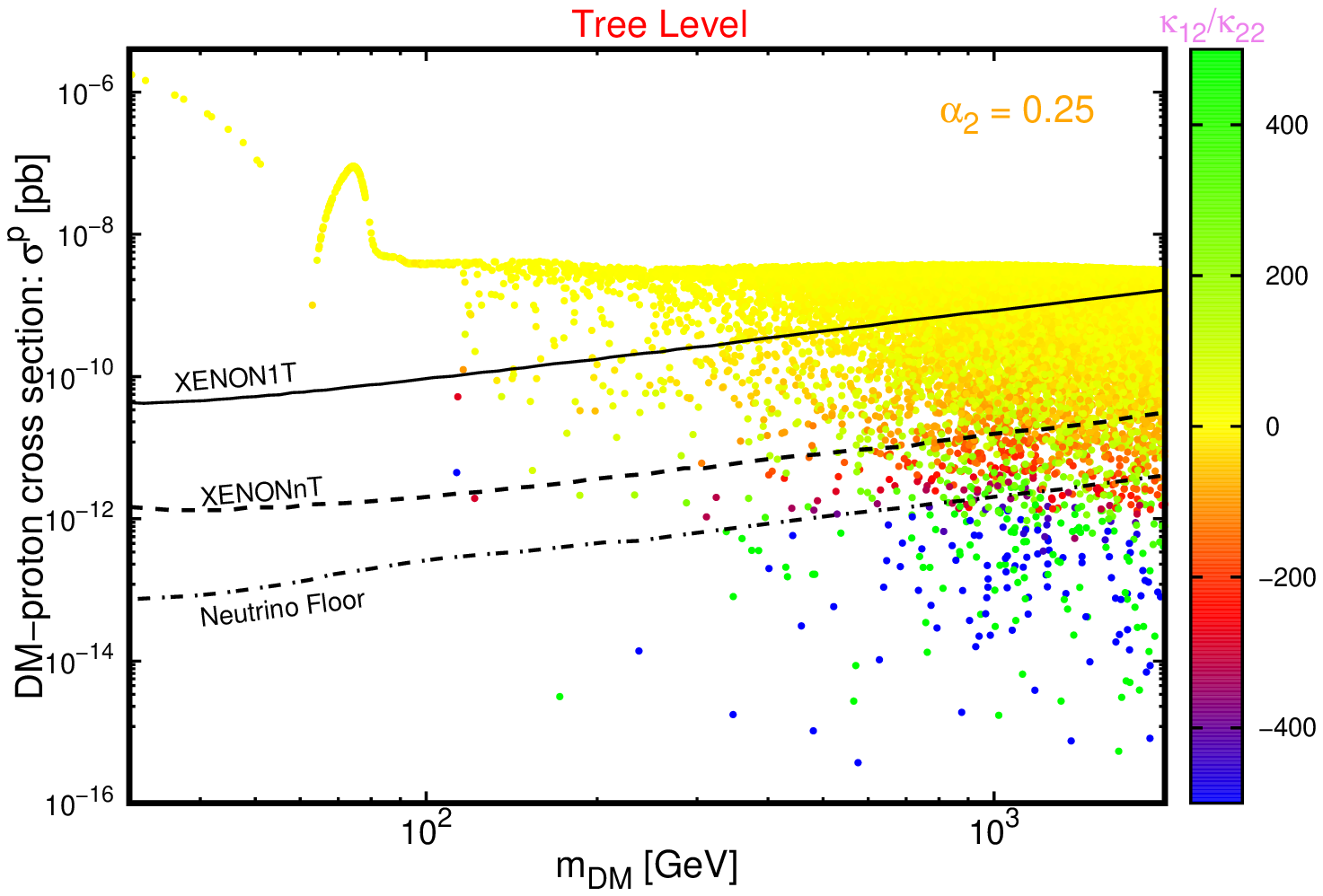}
\end{minipage}
\caption{The DM-nucleon cross section as a function of the DM mass is shown. Upper bounds from XENON1T and XENONnT, and lower bound from neutrino floor are placed. The ratio $\kappa_{11}/\kappa_{22}$ and the ratio $\kappa_{12}/\kappa_{22}$ are shown as color spectrum in the left panel and right panel, respectively. Here $\alpha_2 =0.25$.} 
\label{CrossTree1}
\end{figure}

\begin{figure}
\hspace{-.5cm}
\begin{minipage}{.52\textwidth}
\includegraphics[width=.71\textwidth,angle =-90]{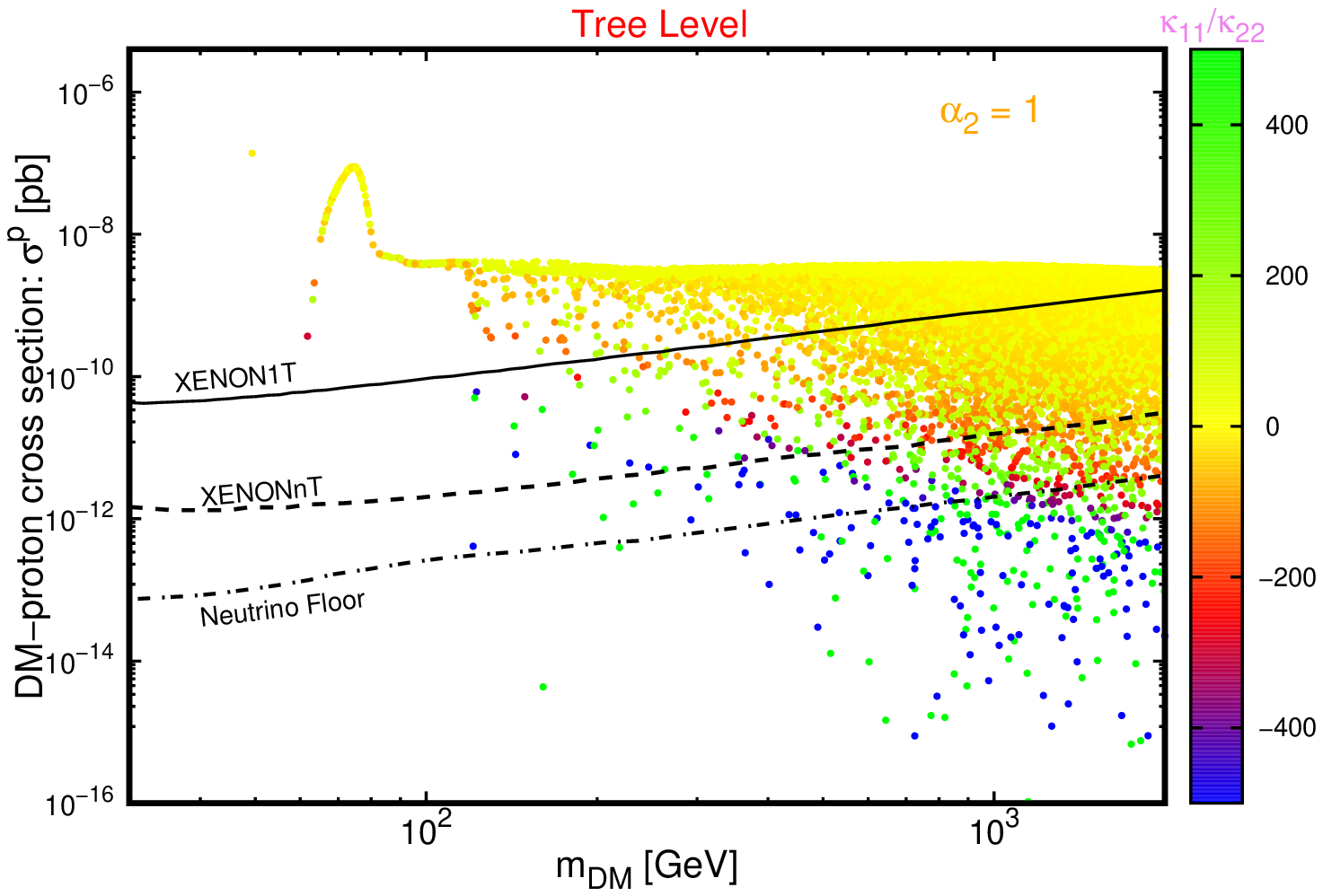}
\end{minipage}
\begin{minipage}{.52\textwidth}
\includegraphics[width=.71\textwidth,angle =-90]{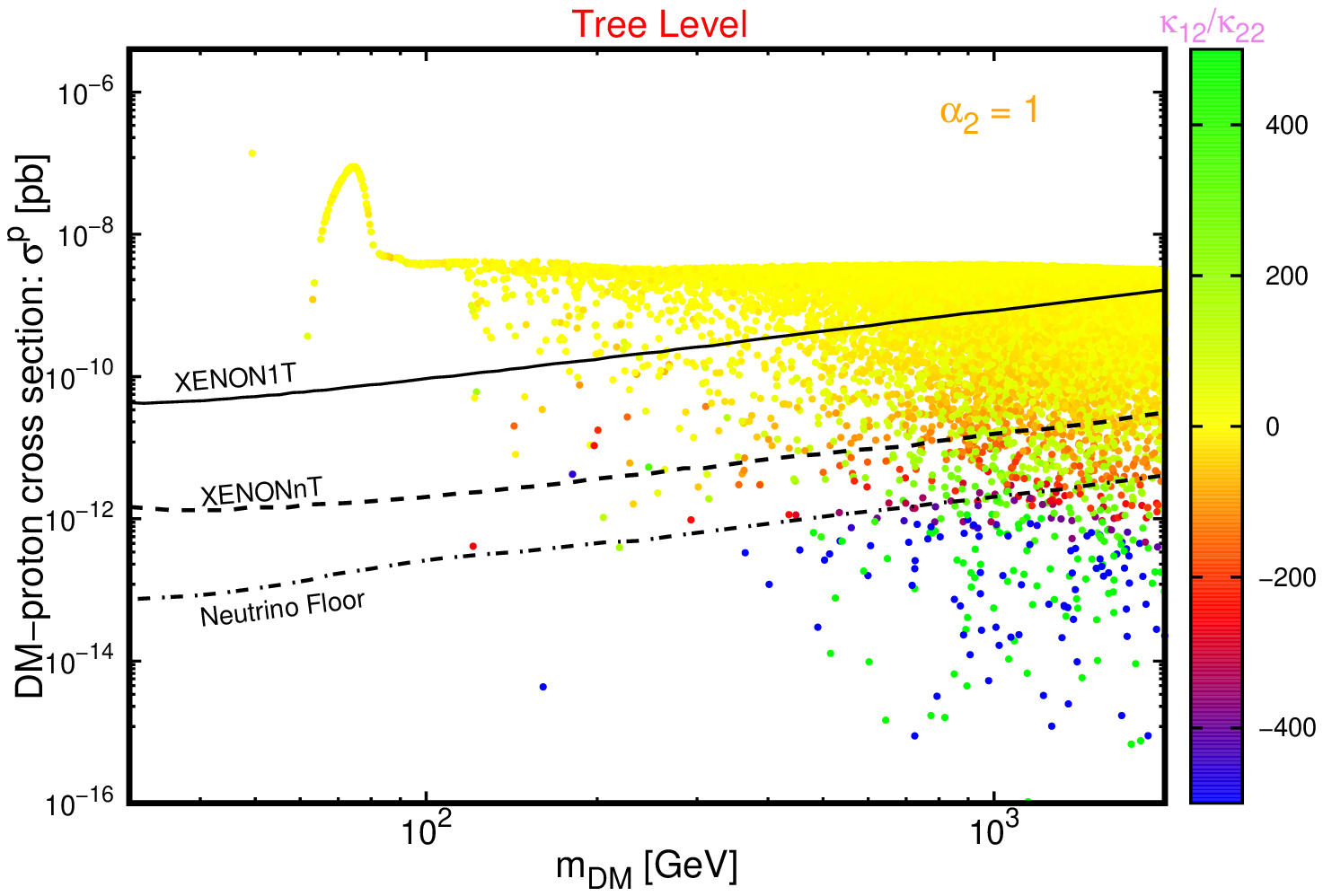}
\end{minipage}
\caption{The same as in Fig.~\ref{CrossTree1} with $\alpha_2 =1$.} 
\label{CrossTree2}
\end{figure}

\begin{figure}
\hspace{-.5cm}
\begin{minipage}{.52\textwidth}
\includegraphics[width=.71\textwidth,angle =-90]{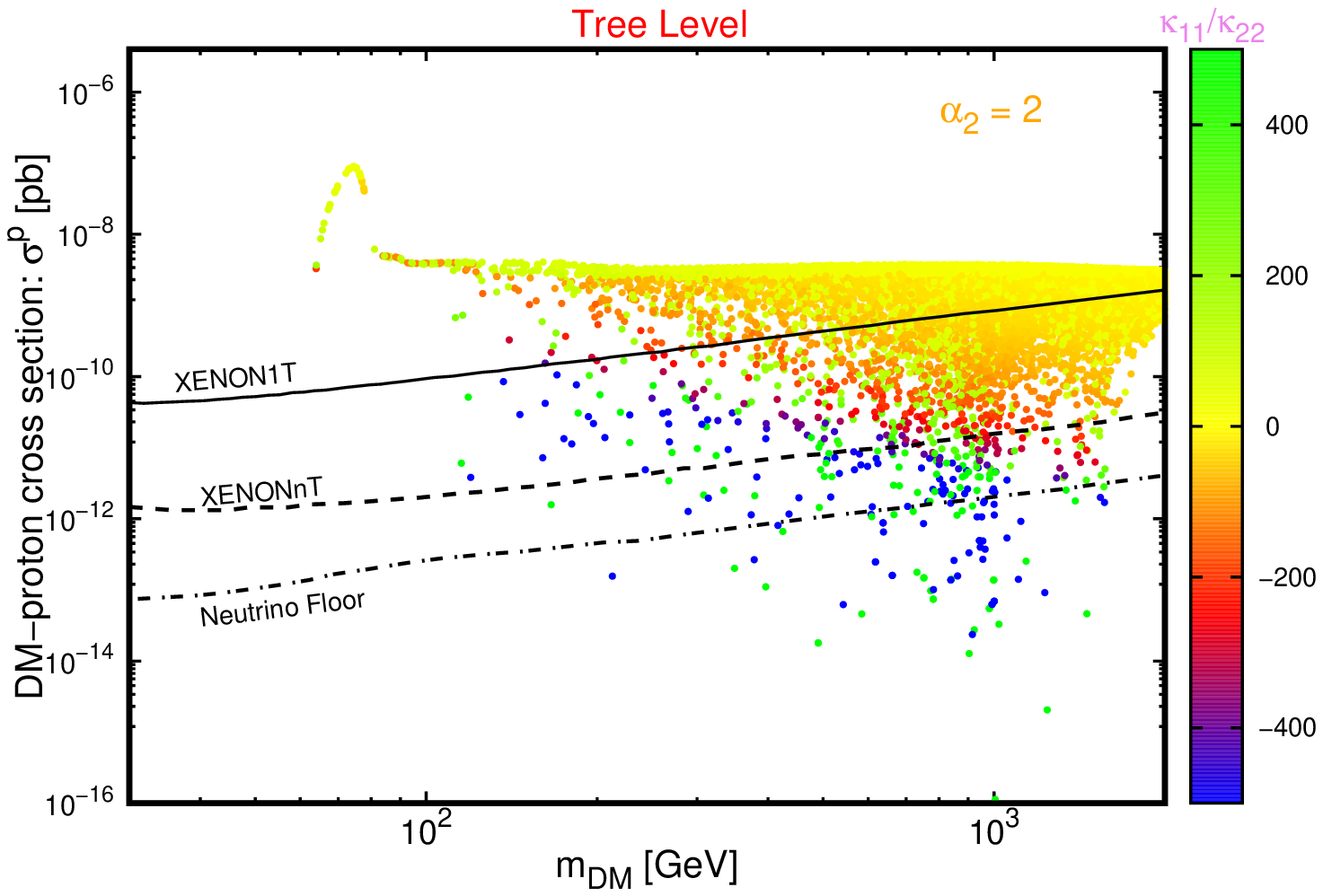}
\end{minipage}
\begin{minipage}{.52\textwidth}
\includegraphics[width=.71\textwidth,angle =-90]{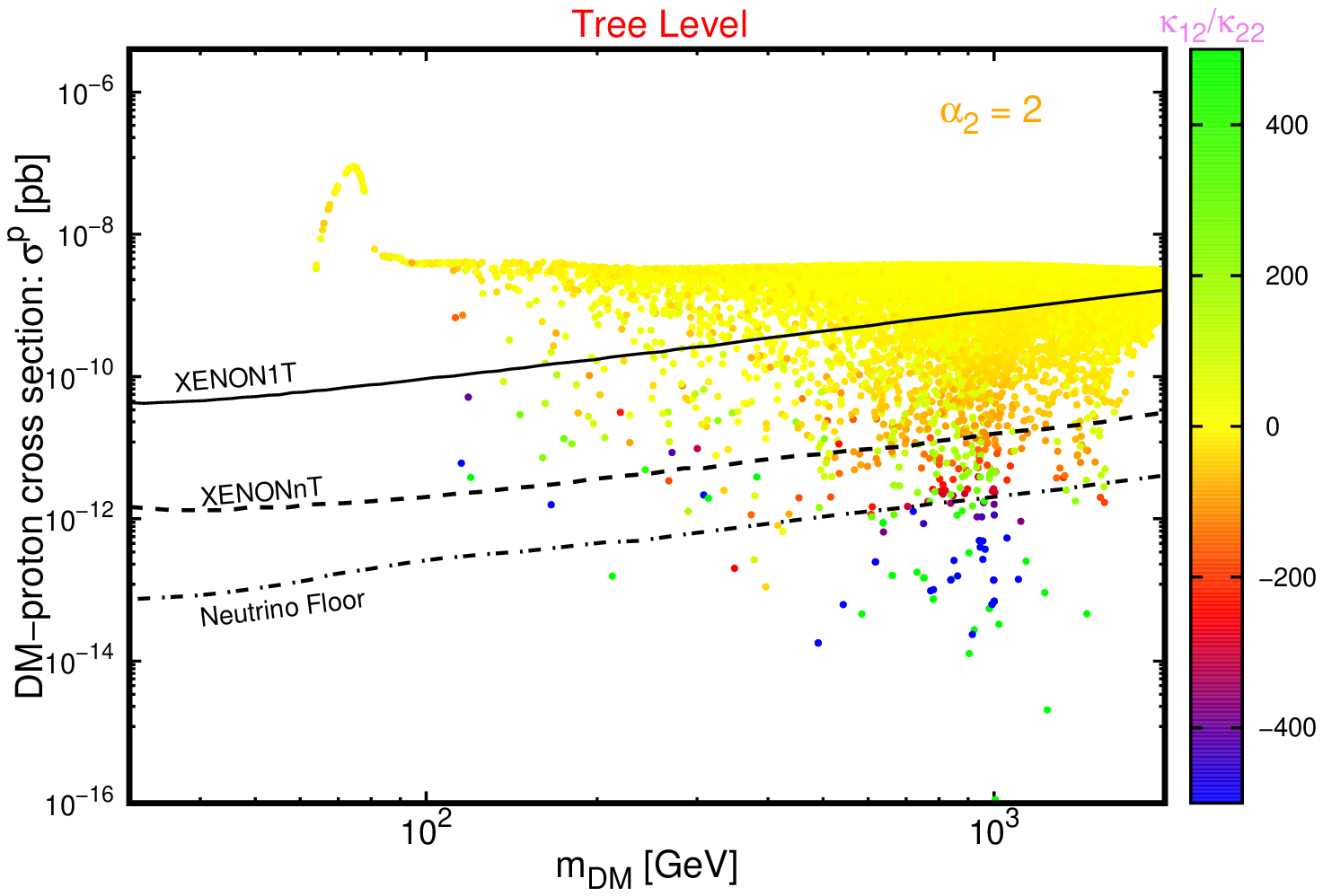}
\end{minipage}
\caption{The same as in Fig.~\ref{CrossTree1} with $\alpha_2 =2$.} 
\label{CrossTree3}
\end{figure}

To proceed further with some numerical results, the model is implemented in the package {\tt MicrOMEGAs} \cite{Alguero:2022inz} to compute 
the relic density and the DM-nucleon cross section at tree level. 
The viable parameter space is found after imposing three different constraints. The upper bounds from XENON1T \cite{XENON:2018voc} and the bound from XENONnT \cite{XENON:2023cxc}.
There is the so-called neutrino floor as a lower limit below which the direct 
detection of dark matter seems scarcely possible \cite{Billard:2021uyg}. 
This bound is also placed to confine the parameter space from below.
We only keep the points in the parameter space where their computed relic density is found 
within the observational range, $\Omega h^2 \sim 0.12$  \cite{Planck:2018vyg}. 
As well, the bound from the invisible Higgs decay has been taken into account. 
The coupling $\alpha_2$ will be fixed at  appropriate values once, $\alpha_2= 0.25, 1, 2$. 
By generating $7\times 10^6$ random points, our scan is performed in the 
following range of the parameters,
\begin{equation}
  0 \le  \alpha_1 \le 1,~~~  0 \le \alpha_{12} \le 2,~~~ 10 \le \delta \le 100~(\text{GeV}),~~~30 \le m_2 \le 2000~(\text{GeV}) \,.
\end{equation}
We present our numerical results for the DM-nucleon cross section 
in terms of the DM mass for $\alpha_2 = 0.25$, $\alpha_2 = 1$ and $\alpha_2 = 2$ in 
Figs.~\ref{CrossTree1}-\ref{CrossTree3}, while all the constraints mentioned above are applied. 
In these plots some new features appear when DM mass exceeds $\sim 125$ GeV. These were absent 
in the singlet scalar model. In fact we realize that regions with DM-nucleon cross section 
below the XENONnT limit and below the neutrino floor open up. 
In these regions the coupling $\kappa_{22}$ picks up small values while
the coupling $\kappa_{12}$ and $\kappa_{11}$ can take large values. As can be seen 
in Fig.~\ref{CrossTree1}, Fig.~\ref{CrossTree2} and Fig.~\ref{CrossTree3}, 
we have $|\kappa_{12}/\kappa_{22}| \gg 1$ and $|\kappa_{11}/\kappa_{22}| \gg 1$. 
This is plausible because now DM annihilation via $t$- and $u$-channel is kinematically allowed, and therefore a large value for $\kappa_{12}$ (and $\kappa_{11}$ in case of co-annihilation) 
regulates the relic density to its right value while 
the coupling $\kappa_{22}$ being a small value has a little impact on the relic density. 
Moreover, we emphasize that the co-annihilation processes are included in our computation for 
relic density, given that based on the standard lore these effects are 
effective when $\delta/m_{\text{DM}} \lesssim 10\%$. 

Why are we encouraged to go beyond tree level in DM-nucleon scattering cross section?
The standard expectation is that the tree level contribution is the dominant part in the cross section. 
According to the results presented in this section we find out that this is not the case for some 
regions in the parameter space. 
For points residing below the XENONnT bound and neutrino floor, we see that 
$|\kappa_{12}/\kappa_{22}| \gg 1$ and $|\kappa_{11}/\kappa_{22}| \gg 1$. 
Therefore, Feynman diagrams at loop level involving only the couplings $\kappa_{12}$ and $\kappa_{11}$, 
might have quite sizable contributions to the cross section in the regions with small DM-nucleon cross section. 
We will dub these loop effects as {\it leading quantum corrections}. 
As we will see in the next section these contributions are only achievable from triangle 
and box diagrams.  

Before we move on further, a comment is appropriate to mention.  
We may think that same type of quantum corrections is involved in the annihilation cross section. In fact what matters here is what we are comparing these corrections to.
In the regions of interest (regions with very small DD cross section) we found  that
$|\kappa_{12}| \gg |\kappa_{22}|$. Looking at the DM annihilation diagrams at tree level 
we see two kind of diagrams: those involving the coupling $\kappa_{22}$ and 
those involving $\kappa_{12}$. In the regions of interest, the annihilation diagrams 
involving $\kappa_{12}$ plays the major role since $|\kappa_{12}| \gg |\kappa_{22}|$.
Now, loop corrections will modify the tiny  coupling $\kappa_{22}$
as $\kappa_{22} \to \kappa_{22} + \delta \kappa_{22}$, where $\delta \kappa_{22}$ comes from loop triangle diagrams involving $\kappa_{12}$. We note here that
$\delta \kappa_{22}$ is much larger than $\kappa_{22}$ but quite smaller than 
$\kappa_{12}$.
Even after this change and working with the effective coupling, the dominant annihilation diagrams 
are still those involving $\kappa_{12}$ because $\kappa_{12}$ is quite 
larger than $\kappa_{22} + \delta \kappa_{22}$. Therefore, the quantum corrections 
in the annihilation diagrams will not change meaningfully the picture we concluded based on our numerical computations. 
On the other hand, in the elastic scattering process we have a tree-level diagram involving $\kappa_{22}$. What we found indicates that the loop diagrams involving 
the effective coupling $\delta \kappa_{22}$ will enhance the DD cross section because 
$\delta \kappa_{22}$ can be quite larger than $\kappa_{22}$.

Moreover, non-zero values of the quartic couplings may add loop correction 
of order $\sim \lambda_{12} \kappa_{11}/(16\pi^2)$ to the 
coupling $\kappa_{22}$  and thus it has effect on the DM-nucleus cross section.  
In this study we assumed that all the quartic couplings are negligible, i.e 
$\lambda_{i} \sim 0$, so the DM-nucleus cross section receives no contribution from 
this type of quantum correction.

\section{Leading Quantum Corrections (LQC)}
\label{LQC}
\begin{figure}
\centering
\includegraphics[width=1\textwidth,angle =0]{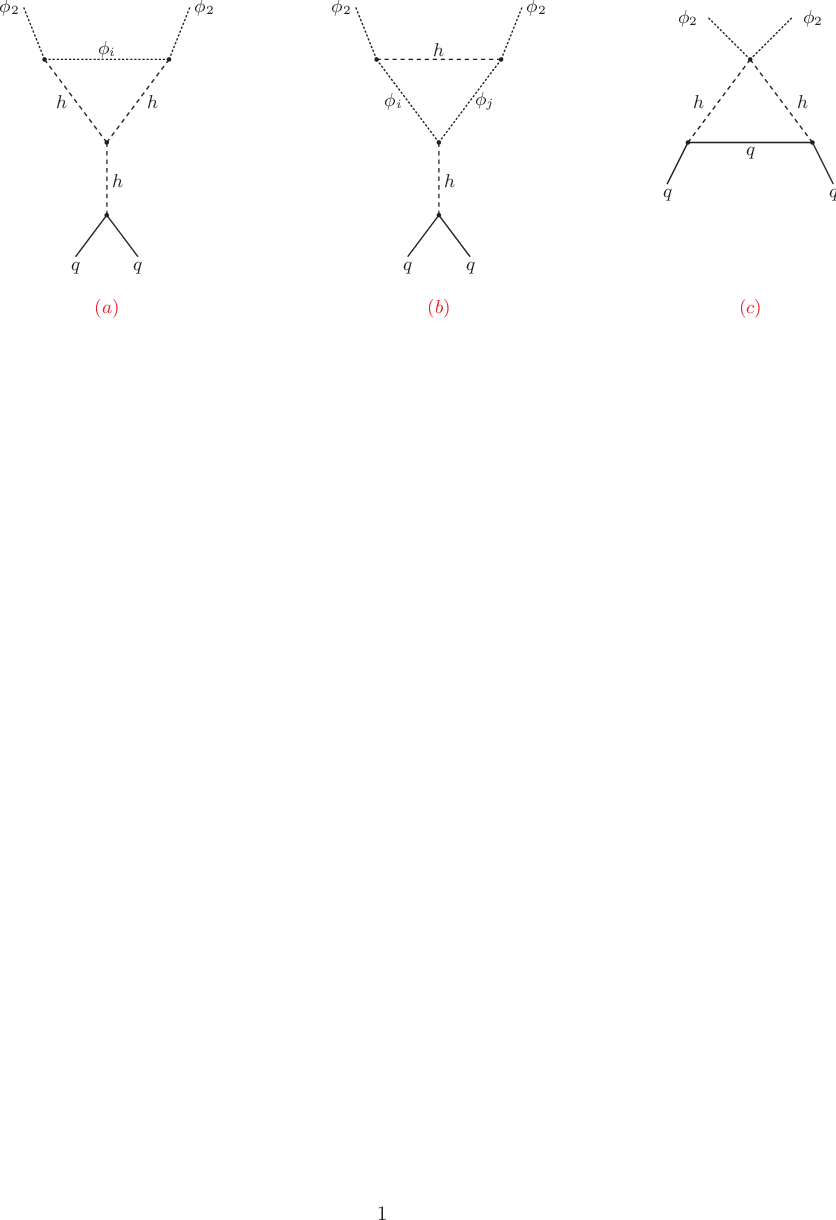}
\caption{Shown are the triangle diagrams for DM scattering off the quarks inside the nucleus. 
Here, $\phi_i = \phi_1, \phi_2$.} 
\label{Triangle-DD}
\end{figure}

\begin{figure}
\centering
\includegraphics[width=.35\textwidth,angle =0]{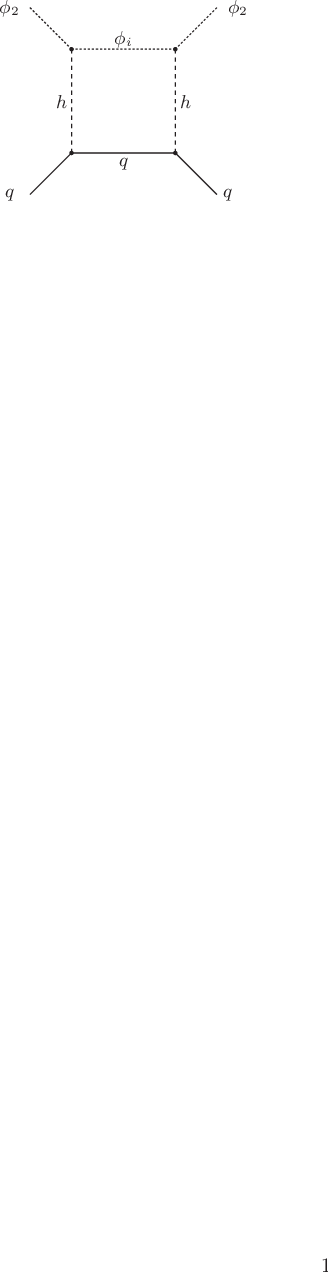}
\caption{The Box diagram for DM scattering off the quarks inside the nucleus is shown. Here, $\phi_i = \phi_1, \phi_2$. } 
\label{Box-DD}
\end{figure}

We present the effective Lagrangian for DM-quark elastic scattering including the 
leading quantum corrections (LQC). These contributions are leading in the sense that they cannot be written as multiplications of a loop factor and the scattering amplitude at tree level. 
Other words, our interest is in those Feynman diagrams at loop level which do not involve the coupling,
 $\kappa_{22}$, the only coupling which enters the DM-nucleon cross section at tree level. 
Since our focus is the regions below the XENONnT limit and below the neutrino floor, 
and since in these regions the coupling $\kappa_{22}$ is 
quite small (while $\kappa_{11}$ and $\kappa_{12}$ are rather large)
therefore at loop level we expect large contributions from diagrams which only involve the couplings $\kappa_{11}$ and $\kappa_{12}$.
There are two types of diagrams which are relevant in this regards; the triangle diagrams 
as shown in Fig.~\ref{Triangle-DD} and a box diagram (both $t$-channel and $u$-channel) represented in Fig.~\ref{Box-DD}. 
Within the triangle and box diagrams we have diagrams that involve the coupling $\kappa_{22}$, however, 
these contributions are relatively very small and are not considered as the leading quantum effects.
The analytical calculations in this section are obtained by employing the Mathematica tool {\tt Package-X} \cite{Patel:2016fam}.

The effective Lagrangian consists of the following parts,
\begin{equation}
 {\cal L}^{\text{LQC}}_{\text{eff}} = 
 \Big( {\cal M}_{(a)}^{\text{Triangle}}  + {\cal M}_{(b)}^{\text{Triangle}} + {\cal M}_{(c)}^{\text{Triangle}}        
    + {\cal M}^{\text{Box}}  \Big)~{\bar q} q  \phi_2 \phi_2  \,.
\end{equation}
The first triangle diagram contains a $\phi_i$ scalar ($\phi_1$ or $\phi_2$), and 
two Higgs particles running in the loop. The effective scattering amplitude at zero momentum transfer 
reads 
\begin{equation}
 {\cal M}_{(a)}^{\text{Triangle}} = \sum_{i=1,2} \frac{m_q}{16\pi^2v m^2_{h}} 4 v^2 \kappa_{i2}^2 c_{hhh} {\cal E}(m^2_2,m^2_i,m^2_h) \,,
\end{equation}
where $c_{hhh} = 3 m^2_{h}/v$, and the loop function ${\cal E}$ is given by   
\begin{equation}
 {\cal E}(m^2_2,m^2_i,m^2_h) = \Big[ \frac{1}{2m_2^2} \log\Big(\frac{m^2_i}{m^2_h}\Big) -
  \frac{m^2_2+m^2_i -m^2_h}{m^2_2\sqrt{\lambda(m^2_2,m^2_h,m^2_i)}} 
  \log\Big( \frac{m^2_h+m^2_i-m^2_2+\sqrt{\lambda(m^2_2,m^2_h,m^2_i)}}{2m_h m_i}  \Big)  \Big] \,,
\end{equation}
where the K{\"a}ll\'{e}n function is given by $\lambda(x,y,z) = x^2+y^2+z^2-2xy-2xz-2yz$.
The amplitude ${\cal M}_{(a)}^{\text{Triangle}}$ consists of two parts. The dominant part is when the scalar
$\phi_1$ runs in the loop, where the resulting amplitude is proportional to $\kappa_{12}^2$. 
The next triangle Feynman diagram with two scalars, $\phi_i$ and $\phi_j$, and a Higgs particle running in the loop gives rise to the following scattering amplitude, 
\begin{equation}
 {\cal M}_{(b)}^{\text{Triangle}} = \sum_{i,j=1,2} \frac{m_q}{16\pi^2v m^2_{h}} 8 v^3 \kappa_{2i}  \kappa_{2j} \kappa_{ij} 
 {\cal F}(m^2_2,m^2_i, m^2_j, m^2_h) ~ {\bar q} q  \phi_2 \phi_2  \,,
\end{equation}
where
\begin{equation}
\begin{split}
 {\cal F}(m^2_2,m^2_i, m^2_j, m^2_h) &=
\Big[ \frac{1}{2m_2^2} \log\Big(\frac{m^2_h}{m^2_j}\Big)  
  + \frac{m^2_h-m^2_2-m^2_i}{2m^2_2(m^2_i-m^2_j)} \log\Big(\frac{m^2_i}{m^2_j}\Big)
\\&
  + \frac{\sqrt{\lambda(m^2_2,m^2_h,m^2_i)}}{m^2_2(m^2_i-m^2_j)} 
    \log\Big(\frac{m^2_h+m^2_i-m^2_2+\sqrt{\lambda(m^2_2,m^2_h,m^2_i)}}{2m_h m_i}  \Big)
\\&
    - \frac{\sqrt{\lambda(m^2_2,m^2_h,m^2_j)}}{m^2_2(m^2_i-m^2_j)} 
    \log\Big(\frac{m^2_h+m^2_j-m^2_2+\sqrt{\lambda(m^2_2,m^2_h,m^2_j)}}{2m_h m_j}  \Big)
    \Big] \,.
\end{split}
\end{equation}
The dominant part of the amplitude is obtained when two identical scalars of type $\phi_1$ runs in the 
loop where the amplitude ${\cal M}_{(b)}^{\text{Triangle}}$ is proportional 
to $\kappa^2_{12} \kappa_{11}$.
In this case we identify ${\cal F}(m^2_2,m^2_i, m^2_i, m^2_h) \equiv {\cal F}(m^2_2,m^2_i, m^2_h)$ 
such that
\begin{equation}
 {\cal F}(m^2_2,m^2_i, m^2_h) = \Big[ \frac{1}{2m_2^2} \log\Big(\frac{m^2_i}{m^2_h}\Big) +
  \frac{m^2_2 - m^2_i + m^2_h}{m^2_2\sqrt{\lambda(m^2_2,m^2_h,m^2_i)}} 
  \log\Big( \frac{m^2_h+m^2_i-m^2_2+\sqrt{\lambda(m^2_2,m^2_h,m^2_i)}}{2m_h m_i}  \Big)  \Big] \,.
\end{equation}
The last triangle diagram has two Higgs and a quark in the loop, with its amplitude as
\begin{equation}
{\cal M}_{(c)}^{\text{Triangle}} =  c_{hh22} (\frac{m_q}{v})^2 {\cal G}(m_h,m_q) \,,
\end{equation}
with the loop function, 
\begin{equation}
\begin{split}
 {\cal G}(m_h,m_q) &= \frac{1}{m_q} + \frac{3m^2_q-m^2_h}{2m^3_q} \log(\frac{m^2_h}{m^2_q})
 \\&
 +\frac{(m^2_h-m^2_q)\sqrt{m^2_h(m^2_h-4m^2_q)}}{m^2_h m^3_q} 
  \log \Big(\frac{m^2_h+\sqrt{m^2_h(m^2_h-4m^2_q)}}{2m_h m_q} \Big) \,,
\end{split}
\end{equation}
and $c_{hh22} = 2 \kappa_{22}$. Since the amplitude is a function of $\kappa_{22}$, numerically 
we expect that it has a very small effect on the cross section in the region of interest.  

The last part of the scattering amplitude comes from box diagram, as shown in Fig.~\ref{Box-DD}. 
The contributions from $t$-channel and $u$-channel are included at zero momentum transfer. 
The effective Lagrangian is obtained by setting the quark mass equal to zero in the 
denominator and we will ignore terms which are momentum suppressed operators generated 
when contracting the quark lines in the numerators. 
The resulting DM-quark interaction is spin-independent. The  final result for 
the effective scattering amplitude is achieved,  
\begin{equation}
 {\cal M}^{\text{Box}} = \sum_{i=1,2} \frac{1}{16\pi^2}(\frac{m_q}{v})^2 v^2 \kappa^2_{i2} 
 m_q \Big[ {\cal H}_1(m_2,m_i,m_h) - {\cal H}_2(m_2,m_i,m_h) \Big] \,,
\end{equation}
where the loop functions ${\cal H}_1$ and ${\cal H}_2$ are provided in Appendix C. 
The part of the amplitude which is proportional to $\kappa^2_{12}$ is dominant. 
Concerning light quarks in the nucleon, the box contribution is suppressed because of the two insertions of the quark-Higgs vertex.      

\section{Numerical Results}
\label{results}
\begin{figure}
\hspace{-.5cm}
\begin{minipage}{.52\textwidth}
\includegraphics[width=.71\textwidth,angle =-90]{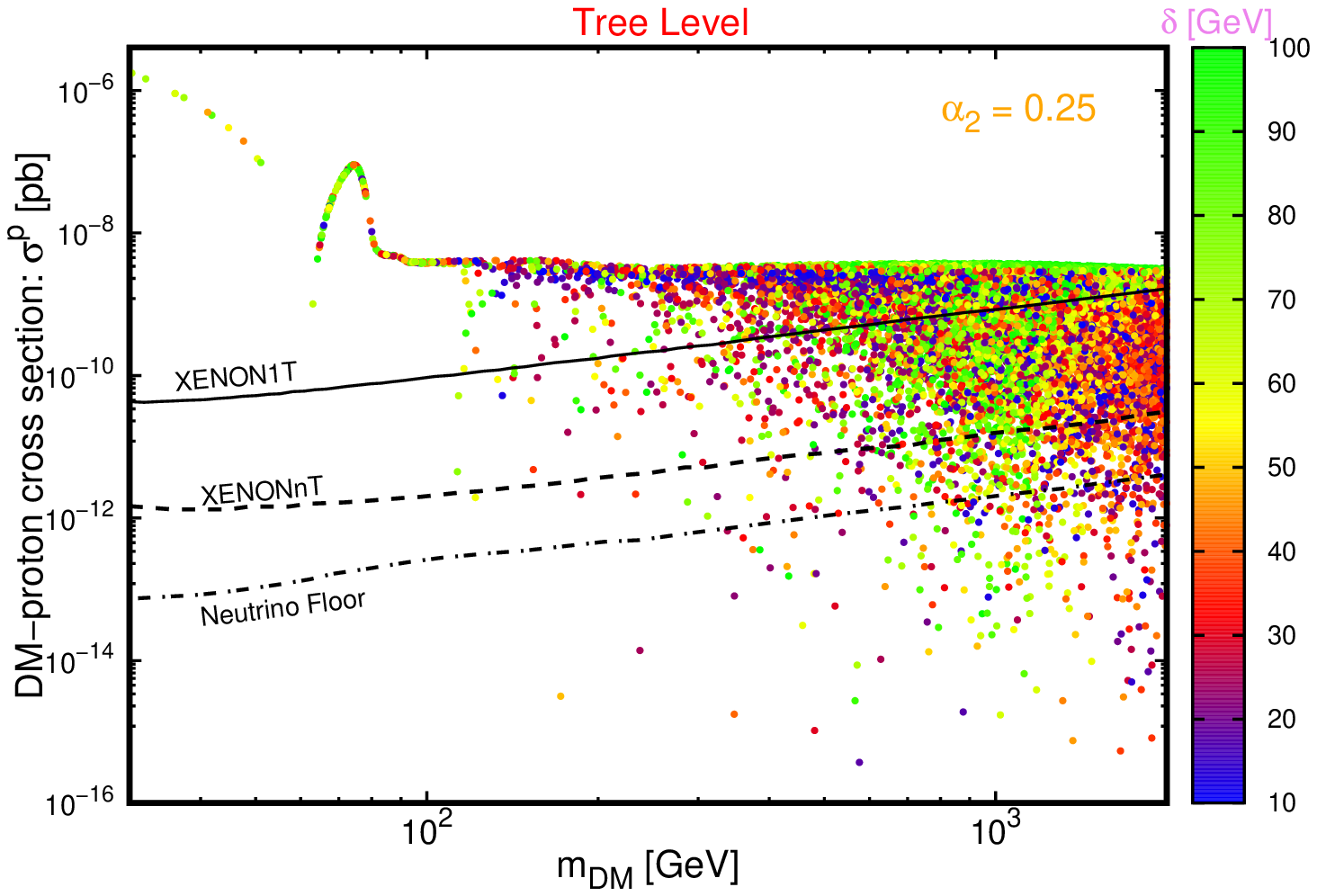}
\end{minipage}
\begin{minipage}{.52\textwidth}
\includegraphics[width=.71\textwidth,angle =-90]{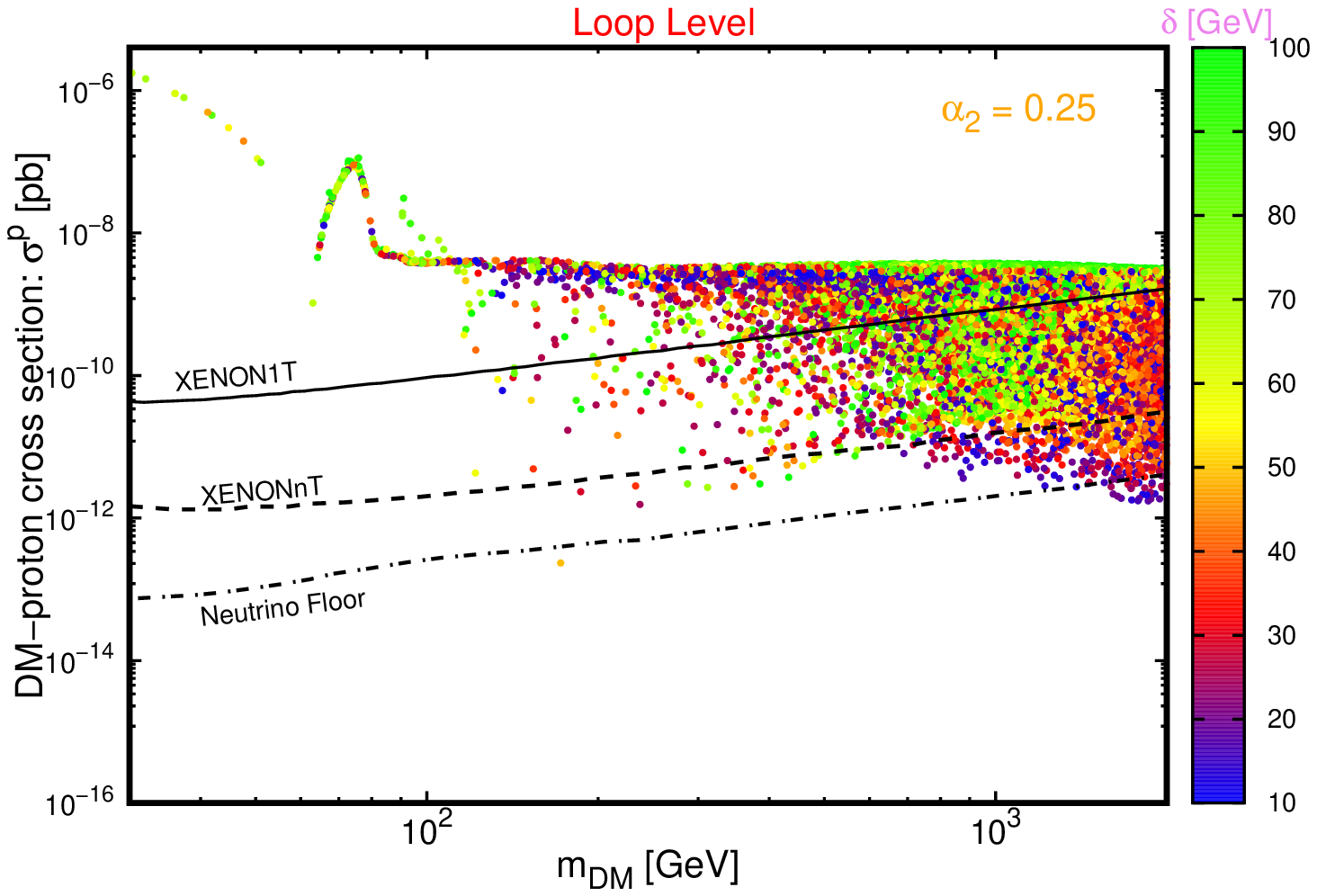}
\end{minipage}
\caption{The DM-nucleon cross section as a function of the DM mass is shown. All the points
respect the observed relic density. The range of the parameter $\delta$ is shown in color spectrum.
Here the free parameter is fixed at $\alpha_2 =0.25$. Bounds from XENON1T, XENONnT and neutrino floor
are placed.} 
\label{LO-NLO-0.25}
\end{figure}

\begin{figure}
\hspace{-.5cm}
\begin{minipage}{.52\textwidth}
\includegraphics[width=.71\textwidth,angle =-90]{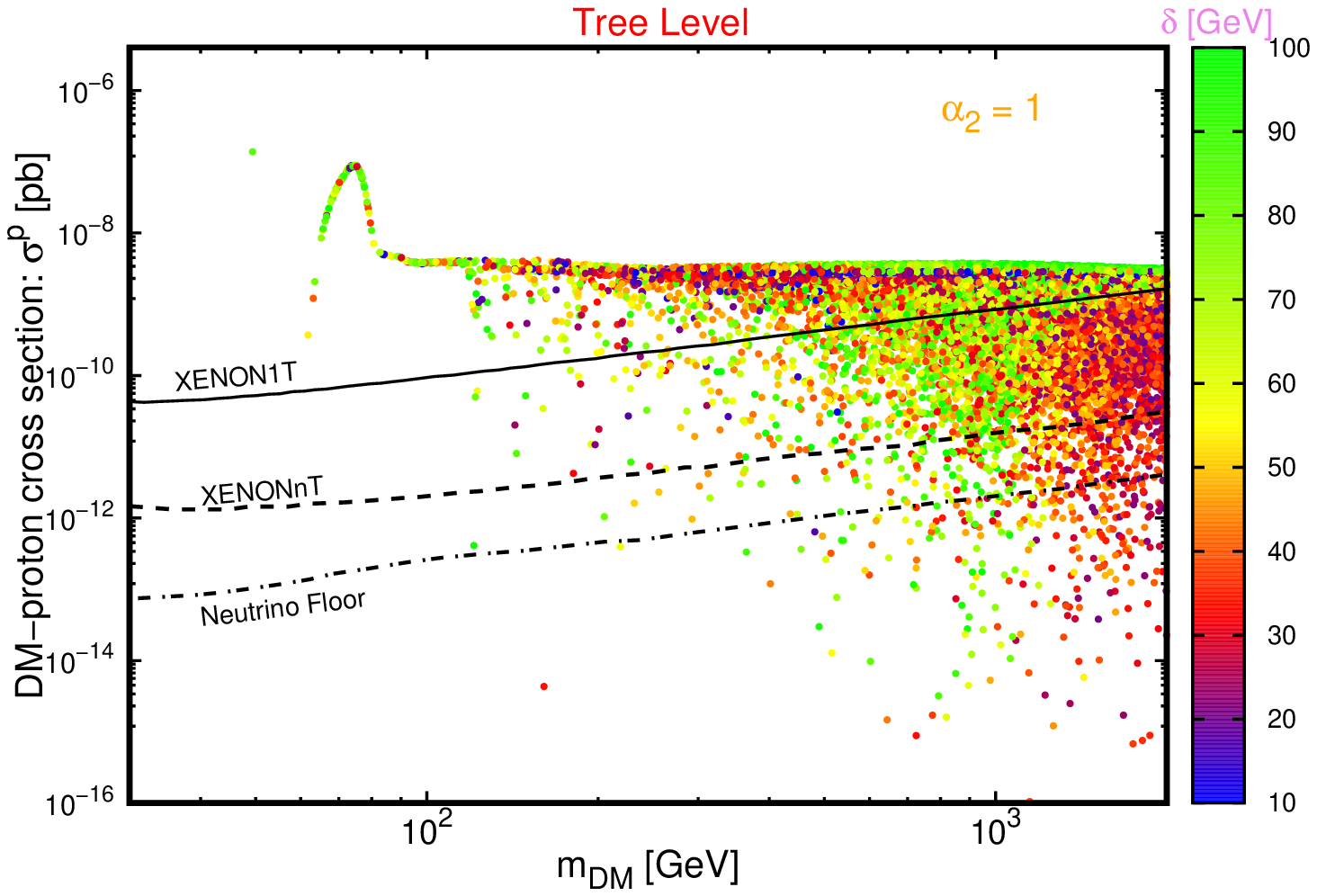}
\end{minipage}
\begin{minipage}{.52\textwidth}
\includegraphics[width=.71\textwidth,angle =-90]{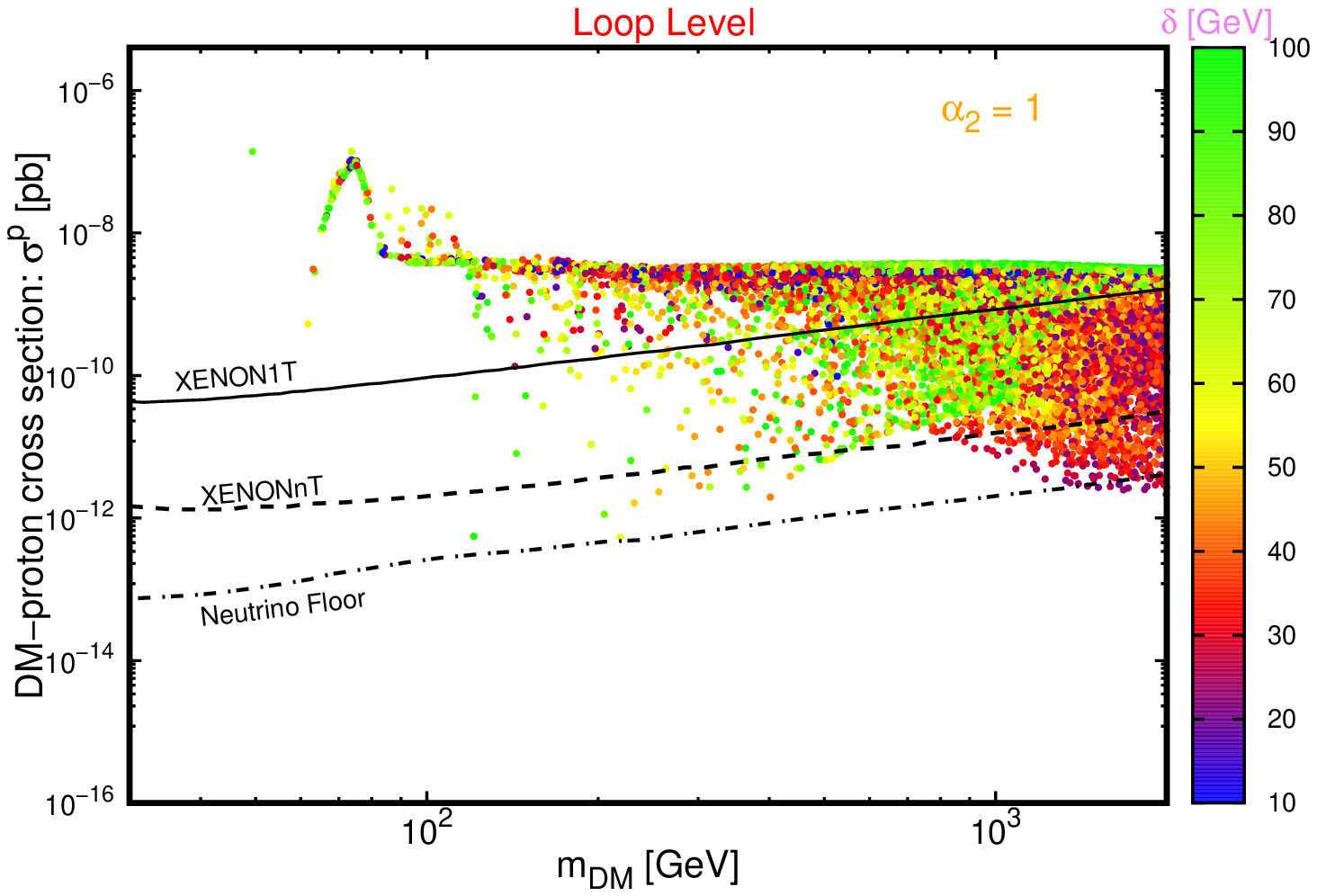}
\end{minipage}
\caption{The same as in Fig.~\ref{LO-NLO-0.25}, with $\alpha_2 =1$.} 
\label{LO-NLO-1}
\end{figure}

\begin{figure}
\hspace{-.5cm}
\begin{minipage}{.52\textwidth}
\includegraphics[width=.71\textwidth,angle =-90]{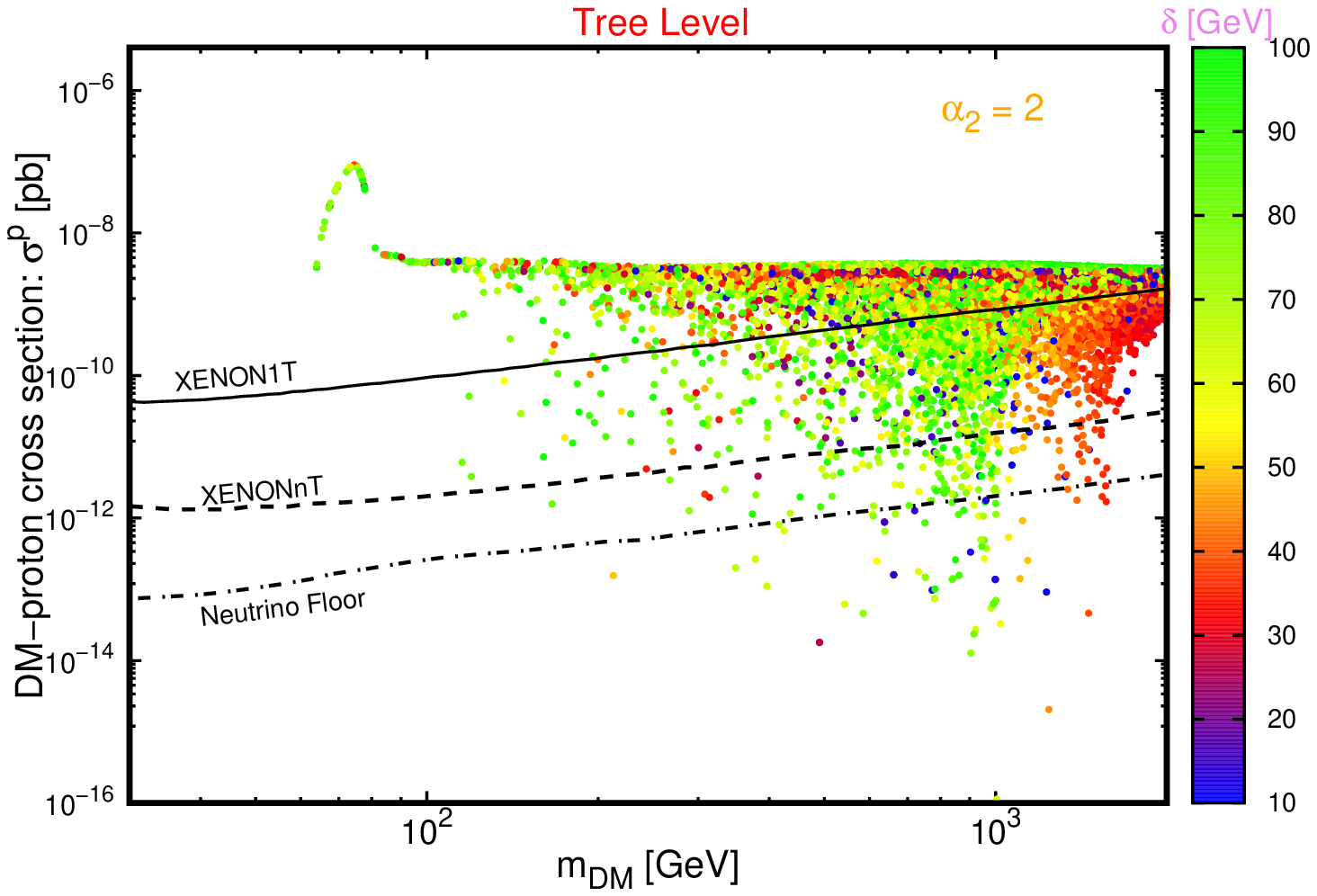}
\end{minipage}
\begin{minipage}{.52\textwidth}
\includegraphics[width=.71\textwidth,angle =-90]{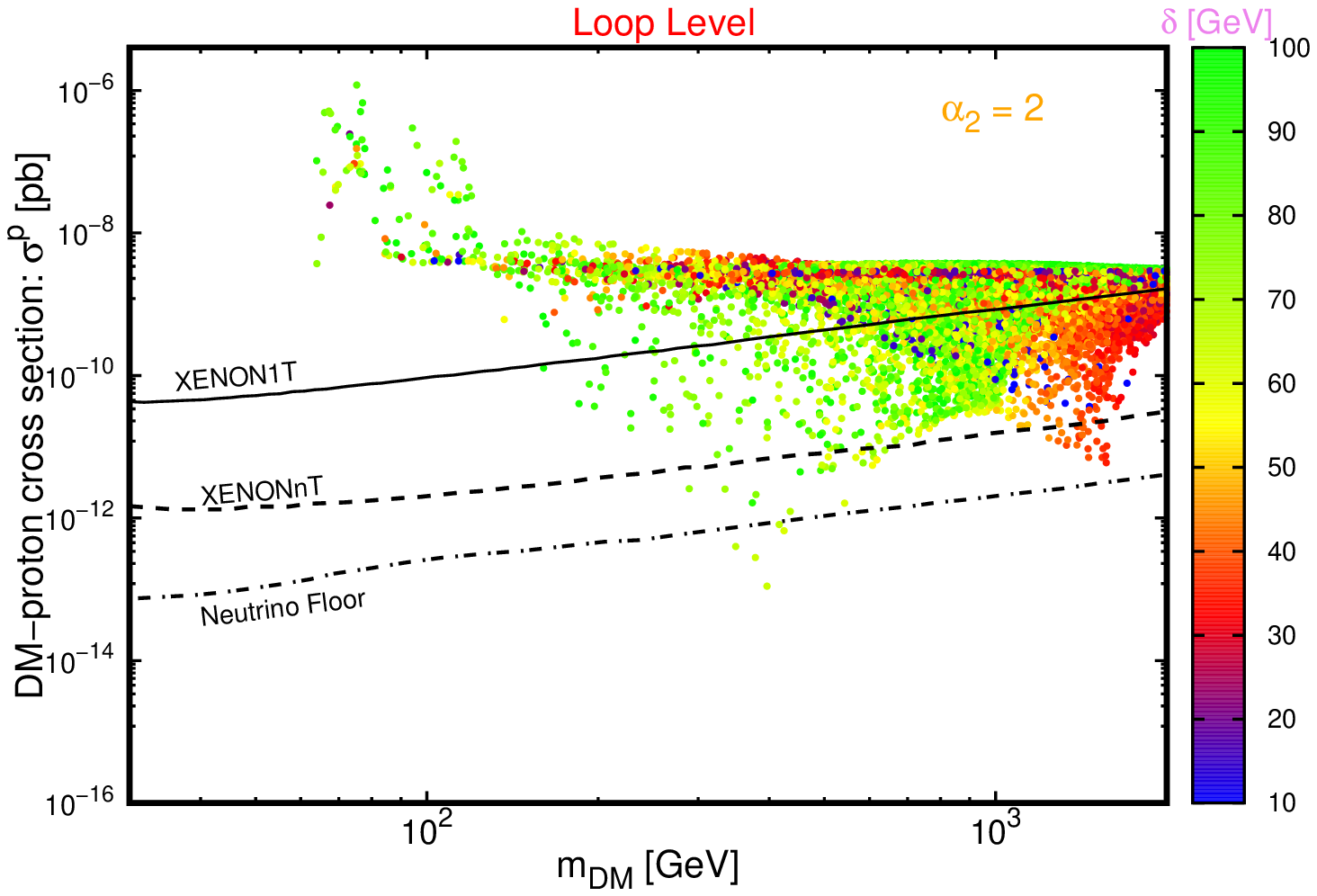}
\end{minipage}
\caption{The same as in Fig.~\ref{LO-NLO-0.25}, with $\alpha_2 =2$.} 
\label{LO-NLO-2}
\end{figure}

In this section we present our main results. 
The task is to continue our numerical calculations by including the 
leading loop corrections in order to find viable regions respecting
the observed relic density and to locate their positions with respect to the upper limits 
from the latest DD experiments and the lower limit from neutrino floor. 
We take the same values for the free parameter $\alpha_2$ as those at tree level, namely $\alpha_2 = 0.25, 1, 2$. The other free parameters are $m_2$,
$\delta$, $\alpha_1$ and $\alpha_{12}$ which vary in the same ranges as set before.
We show the DM-nucleon cross section in terms of the DM mass at tree level and loop level. 
The results are shown for three values of $\alpha_2$ in Fig.~\ref{LO-NLO-0.25}, Fig.~\ref{LO-NLO-1} and Fig.~\ref{LO-NLO-2}. 
Another free parameter shown in the figures is the scalar mass difference, $\delta$.
The findings here comply with what was anticipated about the magnitude of the loop corrections.
In fact we notice the enhancement of the DM-nucleon cross section in regions where the coupling 
$\kappa_{22}$ is quite small in comparison with $\kappa_{11}$ and $\kappa_{12}$, irrespective 
of the size of $\delta$. In case $\alpha_2 = 0.25, 1$ the loop corrections are large 
enough to push almost all the points above the neutrino floor except a small 
patch around $m_\text{DM} \sim 2$ TeV. There are also regions respecting the XENONnT bound
with $m_\text{DM}$ larger than about 100 GeV. 
When $\alpha_2 = 2$, we get a little different feature. There is found a small region
below the neutrino floor centering around $m_\text{DM} = 400$ GeV. In addition there are
two separated regions which respect XENONnT limit; a small 
patch around $m_\text{DM} \sim 1.5$ TeV, and a region in the range $m_\text{DM} \sim 250-600$ GeV.


\section{Conclusion}
\label{conclusion}
While the parameter space of the minimal extension of the SM, i.e. the singlet scalar extended SM, is almost entirely excluded by the current 
direct detection bounds, in two singlet scalar extension of the SM,  a large portion of the parameter space 
can evade the direct detection bounds \cite{Ghorbani:2014gka,Ghorbani:2018hjs}.
In this research, we investigated the effect of the loop corrections on the DM-nucleon cross section 
in two-scalar model to figure out how much these effects may modify the size of 
the viable parameter space.

First in this work we recalculated the relic density, and the DM-nucleon cross section 
at tree level for two-scalar DM scenario, and then have applied the updated bound from DD experiments. 
It is clearly seen that there are not only regions respecting 
the XENONnT bound, but also there exist regions residing below the neutrino floor.

We then defined the leading quantum corrections as those which have the dominant effects
on the regions of interest, namely, points below the XENONnT limit. In these regions 
the coupling $\kappa_{22}$ is quite small while the other two couplings, $\kappa_{11}$ and $\kappa_{12}$
are quite larger. It is only the triangle and box diagrams which can 
bring in sizable contributions to the DM-nucleon cross section, because just in these diagrams 
the couplings $\kappa_{11}$ and $\kappa_{12}$ can appear exclusively. 
To check numerically the impact of the loop effects, we have scanned over the parameter space
taking $m_\text{DM}$, $\delta$, $\alpha_1$ and $\alpha_{12}$ as free parameters while choosing  three discrete values $\alpha_{2} = 0.25, 1, 2$. 
Our main observation is that the leading loop effects are able to shrink the parameter 
space of the two-scalar model at tree level considerably. In other words, the loop quantum effects enlarges the value of the DM-nucleon cross section so that a large part of the parameter space being below the 
neutrino floor is now shifted upward and becomes exploratory regions by the current and 
future DD experiments. This happens independent of the value we choose for the scalar mass 
difference, $\delta$.    

This research confirms the results of other DM models with
large parameter space evading direct detection constraint \cite{Ghorbani:2018pjh,Maleki:2022zuw}, 
in which loop corrections should not be abandoned, particularly for regions 
respecting the DD bounds and regions below the neutrino floor. 

\section{Appendix A}
\label{ApenA} 
As mentioned earlier in the text, the DM annihilation to the SM particles via $s$-channel is possible. 
The corresponding cross section for annihilation to the SM fermions is 
\begin{equation}
\sigma_{\text{ann}} v_{\text{rel}} ( \phi_{2} \phi_{2} \to \bar f f) = 
\frac{N_{c}  m_{f}^2}{\pi} (1-\frac{4m^{2}_{f}}{s})^{\frac{3}{2}}  
\Big[  \frac{\kappa_{22}^2}
{(s-m^{2}_{h})^2+m^{2}_{h}\Gamma^{2}_{h}} \Big] \,,
\end{equation}
and DM annihilation cross section to gauge bosons is 

\begin{equation}
\begin{split}
\sigma_{\text{ann}} v_{\text{rel}} (\phi_{2} \phi_{2} \to W^{+} W^{-},ZZ) &= 
\frac{1}{2\pi s} 
\Big[  \frac{\kappa_{22}^2}
{(s-m^{2}_{h})^2+m^{2}_{h}\Gamma^{2}_{h}} \Big] 
\\&
\times
 \Big[((s-2m_{W}^2)^2+8m_{W}^2)(1-\frac{4m^{2}_{W}}{s})^{\frac{1}{2}} 
\\&
 + \frac{1}{2} ((s-2m_{Z}^2)^2+8m_{W}^2)(1-\frac{4m^{2}_{Z}}{s})^{\frac{1}{2}}  \Big] \,.
\end{split}
\end{equation}
The DM annihilation to the SM Higgs is possible through $s$-, $t$-, $u$-channel and a contact interaction. 
The resulting formula for the cross section is 
\begin{equation}
\begin{split}
 \sigma_{\text{ann}} v_{\text{rel}} ( \phi_{2} \phi_{2} \to hh) &= \frac{\sqrt{1-4m_h^2/s}}{32\pi^2s} \int d\Omega 
\Big[2 \kappa_{22}^2 + \frac{72v^4 \kappa_{22}^2 \lambda_{H}^2}{(s-m_{h}^2)^2}+ 
\frac{v^4\kappa_{12}^4}{(t-m_1^2)^2}
+\frac{v^4\kappa_{12}^4}{(u-m_1^2)^2}
\\&
+\frac{16v^4\kappa_{22}^4}{(t-m_2^2)^2}
+\frac{16v^4\kappa_{22}^4}{(u-m_2^2)^2}
+\frac{16v^2\kappa_{22}^3}{t-m_2^2}
+\frac{16v^2\kappa_{22}}{u-m_2^2}
+\frac{4v^2\kappa_{22} \kappa_{12}^2}{t-m_1^2}
\\&
+\frac{4v^2\kappa_{22} \kappa_{12}^2}{u-m_1^2}
-\frac{24v^2\kappa_{22}^2 \lambda_{H}}{s-m_{h}^2}
-\frac{96v^4\kappa_{22}^3 \lambda_{H}}{(s-m_{h}^2)(t-m_2^2)}
-\frac{96v^4\kappa_{22}^3 \lambda_{H}}{(s-m_{h}^2)(u-m_2^2)}  
\\&
-\frac{24v^4\kappa_{22} \kappa_{12}^2 \lambda_{H}}{(s-m_{h}^2)(t-m_1^2)}
-\frac{24v^4\kappa_{22} \kappa_{12}^2 \lambda_{H}}{(s-m_{h}^2)(u-m_1^2)}  
+\frac{16v^4\kappa_{22}^4}{(t-m_2^2)(u-m_2^2)}
\\&
+\frac{v^4\kappa_{12}^4}{(t-m_1^2)(u-m_1^2)}
+\frac{8v^4 \kappa_{12}^2 \kappa_{22}^2}{(t-m_1^2)(t-m_2^2)}
+\frac{8v^4 \kappa_{12}^2 \kappa_{22}^2}{(t-m_1^2)(u-m_2^2)}
\\&
+\frac{8v^4 \kappa_{12}^2 \kappa_{22}^2}{(u-m_1^2)(t-m_2^2)}
+\frac{8v^4 \kappa_{12}^2 \kappa_{22}^2}{(u-m_1^2)(u-m_2^2)}
\Big] \,,
\end{split}
\end{equation}
where $s$, $t$ and $u$ are the Mandelstam parameters.

\section{Appendix B}
\label{ApenB}
In this section we provide the DM-nucleon cross section at tree level. 
The effective Lagrangian in the limit of negligible momentum transfer induces 
the DM interaction with the quarks inside the nucleon,  
\begin{equation}
 {\cal L}_{\text{eff}} = {\cal C}_q~\phi_2 \phi_2~{\bar q} q  \,,
\end{equation}
where ${\cal C}^{\text{tree}}_q = m_q \kappa_{22}/m^2_h$. 
The elastic scattering cross section of DM-nucleon 
can be obtained by replacing the quark current with nucleon 
current at cost of a proportionality factor. 
The final result we arrive at is a spin-independent (SI) DM-nucleon 
cross section 
\begin{equation}
 \sigma^{\text{N}} = \frac{F^2_N\mu^2_N}{\pi m^2_2} \,,
\end{equation}
in which the parameter $\mu_N$ is the reduced mass of the DM particle and the 
nucleon, and the parameter $F_N$ is connected to the scalar couplings, $f^N$, 
in the following way,
\begin{equation}
\label{form_factor}
 F_N = \sum_{q=u,d,s} \frac{m_N}{m_q} {\cal C}_q f^{N}_q 
 + \frac{2}{27} \sum_{q=c,b,t} \frac{m_N}{m_q} {\cal C}_q f^{N}_g \,, 
\end{equation}
where at tree level, we have
\begin{equation}
 F_N = \Big(\frac{m_N}{m^2_h} \kappa_{22} \Big) f^N  \,.
\end{equation}
Here $m_N$ is the nucleon mass, in our numerical calculations we set it equal to 
the proton mass, and also we take for proton, $f^p \sim 0.284$ \cite{Alguero:2022inz}, 
given that $f^{N}_g = 1- \sum_{q=u,d,s} f^{N}_q$.
The second term in Eq.~\ref{form_factor} arises from effective DM-gluon interactions which 
is obtained based on a relation between $G^{a}_{\mu \nu} G^{a \mu \nu}$ and 
heavy quark current, $\bar Q Q$, \cite{Shifman:1978zn,Abe:2018emu} 
\begin{equation}
 m_Q \bar Q Q  = -\frac{\alpha_s}{12\pi} G^{a}_{\mu \nu} G^{a \mu \nu} \,.
\end{equation}
When we go beyond tree level then contributions from triangle and box 
diagrams (including effective DM-gluon interaction) add extra terms to ${\cal C}_q$ as 
${\cal C}_q = {\cal C}^{\text{tree}}_q + {\cal C}^{\text{triangle}}_q+ {\cal C}^{\text{Box}}_q$.
As pointed out in \cite{Abe:2018emu} a complete treatment of DM-gluon interaction via box diagrams requires full two-loop computations which is beyond the scope of the present work.  

\section{Appendix C}
The explicit expressions for loop functions, ${\cal H}_1$ and ${\cal H}_2$, arising from the box diagram in Fig. \ref{Box-DD} and its $u$-channel counterpart, 
are given as follows,
\begin{equation}
\begin{split}
{\cal H}_1(m_2,m_i,m_h) &=
 \frac{1}{m^2_2 m^2_h} 
 - \frac{3m^2_2+m^2_h-m^2_i}{2m^4_2 m^2_h} \log(\frac{m^2_h}{m^2_i})
  + \frac{3m^4_2+m^4_h-4m^2_2m^2_i-2m^2_h m^2_i+m^4_i}  
    {m^4_2m^2_h\sqrt{\lambda(m^2_2,m^2_h,m^2_i)}}
    \\&
   \times \log \Big(   \frac{m^2_h+m^2_i-m^2_2 + \sqrt{\lambda(m^2_2,m^2_h,m^2_i)}}{2m_h m_i}      \Big)
   - \frac{C(0,0,m^2_2,m_i,0,m_h)}{m^2_2} \,,
\end{split}
\end{equation}
where the scalar function C is 
\begin{equation}
\begin{split}
C(0,0,x,y,0,z) &= 
 -\frac{1}{x}\text{DiLog}\left(\frac{2 \left(x-y^2\right)}{-\sqrt{\lambda \left(x,y^2,z^2\right)}+x-y^2-z^2},x \left(x-y^2\right)\right)
 \\&
 +\frac{1}{x}\text{DiLog}\left(-\frac{2 y^2}{-\sqrt{\lambda \left(x,y^2,z^2\right)}+x-y^2-z^2},-x\right)\\
 & +\frac{1}{x}\text{DiLog}\left(-\frac{2 y^2}{\sqrt{\lambda\left(x,y^2,z^2\right)}+x-y^2-z^2},x\right)
 \\&
 -\frac{1}{x}\text{DiLog}\left(\frac{2 \left(x-y^2\right)}{\sqrt{\lambda \left(x,y^2,z^2\right)}+x-y^2-z^2},-x \left(x-y^2\right)\right)\\
& +\frac{1}{x}\text{Li}_2\left(\frac{y^2-x}{y^2}\right)-\frac{\pi ^2}{6 x} \,.
\end{split}
\end{equation}
and for the function ${\cal H}_2$ we have
\begin{equation}
\begin{split}
{\cal H}_2(m_2,m_i,m_h) = 
      -\frac{1}{m^2_2 m^2_h}
      + \frac{2m^4_2-10 m^2_2 m^2_h -4m^4_h + m^2_2 m^2_i + 5 m^2_h m^2_i -m^4_i}{2m^4_2 m^2_h (2m^2_2+2m^2_h-m^2_i)} \log(\frac{m^2_h}{m^2_i})
  \\
\hspace{-2.cm}  
     + \frac{(2m^2_2 - m^2_i)(m^2_2-3m^2_h+m^2_i)}{m^4_2 m^2_h (2m^2_2+2m^2_h-m^2_i)} 
       \log \Big(\frac{m^2_i}{m^2_i-2m^2_2} \Big)
    \\
\hspace{-.5cm}    
 - \frac{2m^6_2-4m^4_2 m^2_h + 6m^2_2m^4_h-4m^6_h-m^4_2m^2_i+8m^2_2m^2_hm^2_i+9m^4_h m^2_i-2m^2_2 m^4_i-6m^2_hm^4_i+m^6_i}{m^4_2m^2_h(2m^2_2+2m^2_h-m^2_i)\sqrt{\lambda(m^2_2,m^2_h,m^2_i)})}
  \\
  \hspace{-.5cm}
  \times \log \Big( \frac{m^2_h+m^2_i-m^2_2 + \sqrt{\lambda(m^2_2,m^2_h,m^2_i)}}{2m_h m_i}    \Big) 
  -\frac{3D(0,m^2_2,2m^2_2,0,m_h,m_i)}{m^2_2} \,,
\end{split}
\end{equation}
where, the scalar function, D, is 
\begin{equation}
\begin{split}
D(0,x,2x,0,y,z) &=
-\frac{1}{x}\text{DiLog}\left(-\frac{2 x \left(x+2 y^2-z^2\right)}{x \sqrt{\lambda \left(x,y^2,z^2\right)}-x \left(x+3 y^2-z^2\right)},x \left(x+2 y^2-z^2\right)\right)
\\&
+\frac{1}{x}\text{DiLog}\left(-\frac{2 x \left(x+2 y^2-z^2\right)}{-x \sqrt{\lambda \left(x,y^2,z^2\right)}-x \left(x+3 y^2-z^2\right)},-x \left(x+2 y^2-z^2\right)\right)
\\&
-\frac{1}{x}\text{DiLog}\left(-\frac{2 x \left(2 y^2-z^2\right)}{x \sqrt{\lambda \left(x,y^2,z^2\right)}-x \left(x+3 y^2-z^2\right)},x \left(2 y^2-z^2\right)\right)
\\&
-\frac{1}{x}\text{DiLog}\left(-\frac{2 x \left(2 y^2-z^2\right)}{-x \sqrt{\lambda \left(x,y^2,z^2\right)}-x \left(x+3 y^2-z^2\right)},x \left(z^2-2 y^2\right)\right)
\\&
+\frac{1}{x}\text{Li}_2\left(\frac{2 y^2-z^2}{2 y^2}\right)+\frac{1}{x}\text{Li}_2\left(\frac{2 y^2-z^2+2 x}{2 x-z^2}\right)+\frac{1}{x}\text{Li}_2\left(\frac{2 y^2-z^2}{2 y^2-z^2+2 x}\right)
\\&
-\frac{1}{x}\text{Li}_2\left(\frac{2 y^2-z^2+x}{2 y^2-z^2+2 x}\right)+\frac{1}{2 x}\log ^2\left(-\frac{2 y^2}{2 x-z^2}\right) \,.
\end{split}
\end{equation}

\bibliography{ref}
\bibliographystyle{utphys}

\end{document}